\RequirePackage{ifpdf}
\documentclass{JHEP3}
\usepackage{graphicx}  
\usepackage{caption}   
\usepackage{subcaption}
\usepackage{amsmath}   
\usepackage{amssymb}   
\usepackage{feynmp}    
\usepackage{cite}

\def\braket#1{\left\langle #1 \right\rangle}

\def\tr{\textrm{tr}}

\def\cG{\mathcal{G}}
\def\cT{\mathcal{T}}
\def\cQ{\mathcal{Q}}
\def\cA{\mathcal{A}}
\def\cM{\mathcal{M}}
\def\cN{\mathcal{N}}

\def\cZ{\mathcal{Z}}
\def\cW{\mathcal{W}}

\def\IC{\mathbb{C}}
\def\IZ{\mathbb{Z}}
\def\nn{\nonumber}

\def\ap{{\alpha'}}
\def\apt{\ensuremath{\ap \to 0}}

\def\cF{\mathcal F}

\def\Re{\mathrm{Re\,}}
\def\Im{\mathrm{Im\,}}
\def\sg(#1){\textrm{sign}(#1)}
\def\d{\mathrm{d}}

\def\tell{\tilde\ell}
\def\ld{\ell_{(d)}}
\def\tld{\tilde\ell_{(d)}}

\newcommand{\As}{\alpha}         
\newcommand{\Ae}{\beta}          
\newcommand{\At}{\gamma}         

\newcommand{\spbb}[2]{\left[#1#2\right]}
\newcommand{\spaa}[2]{\langle #1 #2\rangle}
\def\BKcite{\cite{Bern:1990cu,Bern:1990ux,Bern:1991aq,Bern:1993wt}}

\title{BCJ duality and double copy in the closed string sector}

\author{Alexander Ochirov, Piotr Tourkine \\
Institut de Physique Th{\'e}orique, CEA-Saclay,
F-91191 Gif-sur-Yvette cedex, France \\
Email: \email{alexander.ochirov@cea.fr}, \email{piotr.tourkine@cea.fr}
}

\abstract{
This paper is focused on the loop-level understanding of the
Bern-Carrasco-Johansson double copy procedure
that relates the integrands of gauge theory and gravity scattering amplitudes.
At four points, the first non-trivial example of that construction is
one-loop amplitudes in $\mathcal N\!=\!2$ super-Yang-Mills theory
and the symmetric realization of $\mathcal N\!=\!4$ matter-coupled supergravity.
Our approach is to use both field and string theory in parallel to analyze these
amplitudes.
The closed string provides a natural framework to analyze the BCJ construction,
in which the left- and right-moving sectors separately create the color and
kinematics at the integrand level.
At tree level, in a five-point example, we show that the
Mafra-Schlotterer-Stieberger procedure gives a new direct proof of the
color-kinematics double copy. We outline the extension of that argument to $n$
points.
At loop level, the field-theoretic BCJ construction of $\mathcal N\!=\!2$ SYM
amplitudes introduces new terms, unexpected from the string theory perspective.
We discuss to what extent we can relate them
to the terms coming from the interactions between left- and right-movers
in the string-theoretic gravity construction.
}
\preprint{IPhT-t13/196}

\begin{document}
\newpage
\section{Introduction}
\label{sec:intro}
The Bern-Carrasco-Johansson color-kinematics duality \cite{Bern:2008qj,Bern:2010ue}
implements in a powerful and elegant way
the relationship between gauge theory and gravity
scattering amplitudes from tree level to high loop orders
\cite{Bern:2008pv,Bern:2009kd,Carrasco:2011mn,Bern:2012uf,
Bern:2011rj,BoucherVeronneau:2011qv,Bern:2012cd,Bern:2012gh,Bern:2013uka,
Carrasco:2012ca,Chiodaroli:2013upa,Huang:2012wr,Boels:2012sy,Boels:2013bi}.
At tree level, this duality is usually perceived in terms of the celebrated
Kawai-Lewellen-Tye relations \cite{Kawai:1985xq},
but a first-principle understanding at loop level is still
missing.\footnote{With the exception of one-loop amplitudes in the self-dual
sector of Yang-Mills theory \cite{Boels:2013bi}.}

In this paper, we search for possible string-theoretic ingredients to understand
the color-kinematics double copy in one-loop four-point amplitudes. 
The traditional ``KLT'' approach, based on the factorization of
closed string \emph{amplitudes} into open string ones: ``open $\times$
open = closed'' at the \emph{integral} level, does not carry over to loop
level.
Instead, one has to look for relations at the \emph{integrand} level. 
In this paper, adopting the approach of \cite{BjerrumBohr:2010zs,Tye:2010dd},
we shall use the fact that the tensor product between the left- and right-moving
sectors of the closed string, i.e. 
\begin{center}
``left-moving $\times$ right-moving = closed'',
\end{center}
relates color and kinematics at the worldsheet integrand level.
\begin{table}[h]
\begin{tabular}{|l|l|l|l|}
  \hline
  \hspace{1pt} Left-moving CFT \hspace{1pt} &
  \hspace{1pt} Right-moving CFT \hspace{1pt} &
  \hspace{1pt} Low-energy limit &
  \hspace{1pt} Closed string theory \\
  \hline
  \hspace{1pt} Spacetime CFT &
  \hspace{1pt} Color CFT &
  \hspace{1pt} Gauge theory & 
  \hspace{1pt} Heterotic \\
  \hspace{1pt} Spacetime CFT &
  \hspace{1pt} Spacetime CFT &
  \hspace{1pt} Gravity theory &
  \hspace{1pt} Type II, (Heterotic) \hspace{1pt}\\
  \hline
\end{tabular}
\caption{Different string theories generating field theories
in the low-energy limit \label{tab:models}}
\end{table}
It is illustrated in table~\ref{tab:models}, where ``Color CFT'' and
``Spacetime CFT'' refer to the respective target-space chiral polarizations and
momenta of the scattered states.
A gauge theory is realized by the closed string when one of the chiral sectors
of the external states is polarized in an internal color space. This is the
basic mechanism of the heterosis which gave rise to the beautiful heterotic
string construction \cite{Gross:1984dd}. 
A gravity theory is realized when both the left- and right-moving polarizations
of the gravitons have their target space in Minkowski spacetime,
as it can be done both in heterotic and type II string. 
In the paper, we shall not describe the gravity sector of the heterotic string,
as it is always non-symmetric. Instead, we will focus on symmetric orbifolds of
the type II string to obtain, in particular, symmetric realizations of
half-maximal ($\cN=4$ in four dimensions) supergravity.

In section~\ref{sec:st-tree}, we review how the closed-string approach
works at tree level with the five-particle example discussed in
\cite{BjerrumBohr:2010zs,Tye:2010dd}. 
We adapt to the closed string the Mafra-Schlotterer-Stieberger procedure
\cite{Mafra:2011kj}, originally used to derive ``BCJ'' numerators in the open
string. 
The mechanism, by which the MSS chiral block representation, in the field theory
limit, produces the BCJ numerators in the heterotic string, works exactly
in the same way in gravity.
However, instead of mixing color and kinematics,
it mixes kinematics with kinematics and results in a form of the amplitude where
the double copy squaring prescription is manifest. We outline a $n$-point proof
of this observation.

Then we thoroughly study the double copy construction in four-point one-loop
amplitudes.
First, we note that the BCJ construction is trivial both in field theory and
string theory when one of the four-point gauge-theory copy corresponds to
$\cN\!=\!4$ SYM.
Then we come to our main subject of study, $\mathcal{N}\!=\!2$ gauge theory and
symmetric realizations of $\mathcal{N}\!=\!4$ gravity amplitudes in four
dimensions. 
We study these theories both in field theory and string theory and compare them
in great detail. 
The real advantage of the closed string in this perspective is that we have
already at hand a technology for building field theory amplitudes from
general string theory models, with various level of supersymmetry and gauge
groups.

In section~\ref{sec:ft-loop}, we provide a BCJ construction of half-maximal
supergravity coupled to matter fields as a double copy of $\cN=2$ SYM.
Then in section~\ref{sec:st-loop}, we give the string-based integrands and
verify that they integrate to the same gauge theory and gravity amplitudes.
Finally, we compare the two calculations in section~\ref{sec:comparison} by
transforming the field-theoretic loop-momentum expressions to the same
worldline form as the string-based integrands, and try to relate the BCJ
construction to the string-theoretic one.

Both of them contain box diagrams,
but the field-theoretic BCJ construction of gauge theory amplitudes
has additional triangles, which integrate to zero
and are invisible in the string-theoretic derivation.
Interestingly, at the integrand level,
the comparison between the BCJ and the string-based boxes
is possible only up to a new total derivative term,
which we interpret as the messenger of the BCJ representation information in
the string-based integrand. 
However, we argue that, against expectations,
this change of representation cannot be obtained by integrations by part,
and we suggest that this might be linked to our choice of the BCJ
representation. Therefore, it provides non-trivial physical information on the
various choices of BCJ ansatzes.

The square of the BCJ triangles later contributes to the gravity amplitude.
String theory also produces a new term on the gravity side, which is due to
left-right contractions.
We manage to relate it to triangles squared and parity-odd terms squared,
which is possible up to the presence of ``square-correcting-terms'', whose
appearance we argue to be inevitable and of the same dimensional nature as the
string-theoretic left-right contractions.

We believe that our work constitutes a step towards a string-theoretic
understanding of the double copy construction at loop level in theories with
reduced supersymmetry, although some facts remain unclarified. For
instance, it seems that simple integration-by-part identities are not enough to
obtain some BCJ representations (e.g.~ours) from string theory.

\section{Review of the BCJ construction}
\label{sec:review}

In this section, we briefly review the BCJ duality and the double copy construction
in field theory, as well as the current string-theoretic understanding of these issues
(see also the recent review \cite[section~13]{Elvang:2013cua}).

To begin with, consider a $n$-point $L$-loop color-dressed amplitude in gauge theory
as a sum of Feynman diagrams.
The color factors of graphs with quartic gluon vertices,
written in terms of the structure constants $f^{abc}$,
can be immediately understood as sums of cubic color diagrams.
Their kinematic decorations can also be adjusted, in a non-unique way,
so that their pole structure would correspond to that of trivalent diagrams.
This can be achieved
by multiplying and dividing terms by the denominators of missing propagators.
Each four-point vertex can thus be interpreted as a $s$-, $t$- or $u$-channel tree,
or a linear combination of those.
By performing this ambiguous diagram-reabsorption procedure,
one can represent the amplitude as a sum of cubic graphs only:
      \begin{equation} \begin{aligned}
            \mathcal{A}_n^L = i^L g^{n+2L-2}
                  \!\!\!\!\!\!\!
                  \sum_{\text{cubic graphs} \; \Gamma_i}
                  \int \prod_{j=1}^{L} \frac{\d^d \ell_j}{(2\pi)^d}
                  \frac{1}{S_i}
                  \frac{c_i \, n_i(\ell)}{D_i(\ell)} \,,
      \label{e:Ageneral}
	\end{aligned} \end{equation}
where the denominators $D_i$, symmetry factors $S_i$ and color factors $c_i$
are understood in terms of the Feynman rules of the adjoint scalar $\phi^3$-theory
(without factors of $i$)
and the numerators $n_i$ generically lose their Feynman rules interpretation.

      \begin{figure}[t]
      \centering
      \parbox{60pt}{ \begin{fmffile}{ttree}
      \fmfframe(10,10)(10,0){ \begin{fmfgraph*}(40,40)
            \fmflabel{$1$}{g1}
            \fmflabel{$2$}{g2}
            \fmflabel{$3$}{g3}
            \fmflabel{$4$}{g4}
            \fmfleft{g1,g2}
            \fmfright{g4,g3}
            \fmf{plain}{g1,v14,g4}
            \fmf{plain}{g2,v23,g3}
            \fmf{plain,tension=0.70}{v14,v23}
      \end{fmfgraph*} }
      \end{fmffile} }
      $ - $
      \parbox{60pt}{ \begin{fmffile}{utree}
      \fmfframe(10,10)(10,0){ \begin{fmfgraph*}(40,40)
            \fmflabel{$1$}{g1}
            \fmflabel{$2$}{g2}
            \fmflabel{$4$}{g4}
            \fmflabel{$3$}{g3}
            \fmfleft{g1,g2}
            \fmfright{g3,g4}
            \fmf{plain}{g1,v13,g3}
            \fmf{plain}{g2,v24,g4}
            \fmf{plain,tension=0.70}{v13,v24}
      \end{fmfgraph*} }
      \end{fmffile} }
      $ = $
      \parbox{60pt}{ \begin{fmffile}{stree}
      \fmfframe(10,10)(10,0){ \begin{fmfgraph*}(60,40)
            \fmflabel{$1$}{g1}
            \fmflabel{$2$}{g2}
            \fmflabel{$3$}{g3}
            \fmflabel{$4$}{g4}
            \fmfleft{g1,g2}
            \fmfright{g4,g3}
            \fmf{plain}{g1,v12,g2}
            \fmf{plain}{g3,v34,g4}
            \fmf{plain,tension=0.70}{v12,v34}
      \end{fmfgraph*} }
      \end{fmffile} } \\
      \hspace{16pt}$c_t \text{ or } n_t$
      \hspace{35pt}$c_u \text{ or } n_u$
      \hspace{44pt}$c_s \text{ or } n_s$
      \caption{Basic Jacobi identity \label{fig:jacobi0} for the color factors.}
      \end{figure}
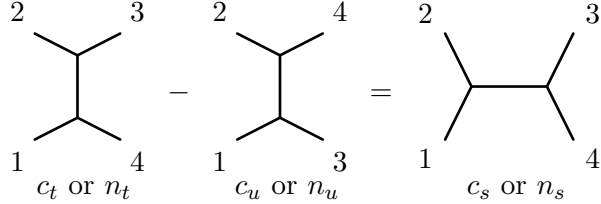

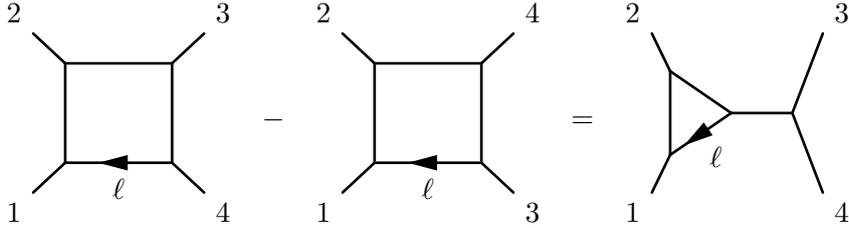
\begin{figure}[t]
      \centering
      \parbox{100pt}{ \begin{fmffile}{graph7a}
      \fmfframe(10,10)(10,0){ \begin{fmfgraph*}(80,60)
            \fmflabel{$1$}{g1}
            \fmflabel{$2$}{g2}
            \fmflabel{$3$}{g3}
            \fmflabel{$4$}{g4}
            \fmfleft{g1,g2}
            \fmfright{g4,g3}
            \fmf{plain}{g1,v1}
            \fmf{plain}{g2,v2}
            \fmf{plain}{g3,v3}
            \fmf{plain}{g4,v4}
            \fmf{plain,tension=0.30}{v1,v2}
            \fmf{plain,tension=0.30}{v2,v3}
            \fmf{plain,tension=0.30}{v3,v4}
            \fmf{plain_arrow,tension=0.30,label=$\ell$,
                                          label.side=left}{v4,v1}
      \end{fmfgraph*} }
      \end{fmffile} }
      $ - $
      \parbox{100pt}{ \begin{fmffile}{graph8a}
      \fmfframe(10,10)(10,0){ \begin{fmfgraph*}(80,60)
            \fmflabel{$1$}{g1}
            \fmflabel{$2$}{g2}
            \fmflabel{$4$}{g4}
            \fmflabel{$3$}{g3}
            \fmfleft{g1,g2}
            \fmfright{g3,g4}
            \fmf{plain}{g1,v1}
            \fmf{plain}{g2,v2}
            \fmf{plain}{g3,v3}
            \fmf{plain}{g4,v4}
            \fmf{plain_arrow,tension=0.30,label=$\ell$,
                                          label.side=left}{v3,v1}
            \fmf{plain,tension=0.30}{v1,v2}
            \fmf{plain,tension=0.30}{v2,v4}
            \fmf{plain,tension=0.30}{v4,v3}
      \end{fmfgraph*} }
      \end{fmffile} }
      $ = $
      \parbox{100pt}{ \begin{fmffile}{graph6a}
      \fmfframe(10,10)(10,0){ \begin{fmfgraph*}(80,60)
            \fmflabel{$1$}{g1}
            \fmflabel{$2$}{g2}
            \fmflabel{$3$}{g3}
            \fmflabel{$4$}{g4}
            \fmfleft{g1,g2}
            \fmfright{g4,g3}
            \fmf{plain}{g1,v1}
            \fmf{plain}{g2,v2}
            \fmf{plain,tension=0.60}{g3,v34,g4}
            \fmf{plain,tension=0.60}{v12,v34}
            \fmf{plain,tension=0.30}{v1,v2}
            \fmf{plain,tension=0.30}{v2,v12}
            \fmf{plain_arrow,tension=0.30,label=$\ell$,
                                          label.side=left}{v12,v1}
      \end{fmfgraph*} }
      \end{fmffile} }
      \caption{Sample Jacobi identity for one-loop numerators \label{fig:jacobi1}}
      \end{figure}

Note that the antisymmetry $f^{abc}=-f^{bac}$ and the Jacobi identity
\begin{equation}
      f^{a_2 a_3 b} f^{b a_4 a_1}
    - f^{a_2 a_4 b} f^{b a_3 a_1} =  f^{a_1 a_2 b} f^{b a_3 a_4} \,,
\label{e:jacobi0}
\end{equation}
shown pictorially in figure~\ref{fig:jacobi0}
induces numerous algebraic relations among the color factors,
such as the one depicted in figure~\ref{fig:jacobi1}.

We are now ready to introduce the main constraint of
\emph{the BCJ color-kinematics duality} \cite{Bern:2008qj,Bern:2010ue}:
let the kinematic numerators $n_i$, defined so far very vaguely,
satisfy the same algebraic identities as their corresponding color factors $c_i$:
\begin{equation} \begin{aligned}
      c_i =-c_j ~~~&\Leftrightarrow~~~ n_i = -n_j  \,, \\
      c_i - c_j = c_k ~~~&\Leftrightarrow~~~ n_i - n_j = n_k \,.
\label{e:duality}
\end{aligned} \end{equation}
This reduces the freedom in the definition of $\{n_i\}$ substantially, but not entirely,
to the so-called generalized gauge freedom.
The numerators that obey the duality \ref{e:duality} are called the BCJ numerators.
Note that even the basic Jacobi identity~\eqref{e:jacobi0},
obviously true for the four-point tree-level color factors,
is much less trivial when written for the corresponding kinematic numerators.

Once imposed for gauge theory amplitudes, that duality results in
\emph{the BCJ double copy} construction for gravity amplitudes
in the following form:\footnote{In the rest of this paper,
we omit trivial coupling constants by setting $g=1$, $\kappa=2$.
At one loop, we can also rescale the numerators by a factor of $-i$
to completely eliminate the prefactors in \eqref{e:Ageneral} and
\eqref{e:Mgeneral}.}
      \begin{equation} \begin{aligned}
            \mathcal{M}_n^L = i^{L+1} \left( \frac{\kappa}{2} \right)^{n+2L-2}
                  \!\!\!\!\!\!\!
                  \sum_{\text{cubic graphs} \; \Gamma_i}
                  \int \prod_{j=1}^{L} \frac{\d^d \ell_j}{(2\pi)^d}
                  \frac{1}{S_i}
                  \frac{n_i(\ell) \, \tilde{n}_i(\ell)}{D_i(\ell)} \,,
      \label{e:Mgeneral}
	\end{aligned} \end{equation}
where only one of the numerator sets, $\{n_i\}$ or $\{\tilde{n}_i\}$,
needs to obey the color-kinematics duality \eqref{e:duality}.
In this way, gauge and gravity theories are related
at the integrand level in loop momentum space.
In this paper, we loosely refer to eqs.~\eqref{e:Ageneral} and \eqref{e:Mgeneral},
related by the duality \eqref{e:duality}, as \emph{the BCJ construction}.

      A comment is due at loop level:
the loop-momentum dependence of numerators $n_i(\ell)$ should be traced with care.
For instance, in the kinematic Jacobi identity given in figure~\ref{fig:jacobi1},
one permutes the legs $3$ and $4$, but keeps the momentum $\ell$ fixed,
because it is external to the permutation.
Indeed, if one writes that identity for the respective color factors,
the internal line $\ell$ will correspond to the color index
outside of the basic Jacobi identity of figure~\ref{fig:jacobi0}.
In general, the correct loop-level numerator identities
correspond to those for the unsummed color factors
in which the internal-line indices are left uncontracted.

Formulas \eqref{e:Ageneral} and \eqref{e:Mgeneral}
are a natural generalization of the original discovery at tree level
\cite{Bern:2008qj}.
The double copy for gravity \eqref{e:Mgeneral} has been proven in
\cite{Bern:2010yg} to hold to any loop order,
if there exists a BCJ representation \eqref{e:Ageneral}
for at least one of the gauge theory copies.
Such representations were found in numerous calculations
\cite{Bern:2010ue,Carrasco:2011mn,Bern:2012uf,Vanhove:2010nf,
Bern:2011rj,BoucherVeronneau:2011qv,Bern:2012cd,Bern:2012gh,Bern:2013uka,
Carrasco:2012ca,Chiodaroli:2013upa, Huang:2012wr,Bern:2013qca}
up to four loops in $\cN=4$ SYM \cite{Bern:2009kd}.
A systematic way to find BCJ numerators is known
for Yang-Mills theory at tree level \cite{Tolotti:2013caa},
and in $\cN=4$ SYM at one loop \cite{Bjerrum-Bohr:2013iza}.
Moreover, for a restricted class of amplitudes
in the self-dual sectors of gauge theory and gravity,
one can trace the Lagrangian origin of the infinite-dimensional kinematic Lie algebra
\cite{Monteiro:2011pc,Boels:2013bi}.

The string-theoretic understanding of the double copy at tree level
dates back to the celebrated KLT relations \cite{Kawai:1985xq}
between tree-level amplitudes in open and closed string theory,
later improved with the discovery of monodromy relations and the momentum kernel
in
\cite{Stieberger:2009hq,BjerrumBohr:2009rd,Tye:2010dd,BjerrumBohr:2010zs,
BjerrumBohr:2010hn}.
In the field theory limit, these relations implement the fact that
in amplitudes the degrees of freedom of a graviton can be
split off into those of two gauge bosons.
Recently, a new chiral block representation of the open-string integrands
was introduced~\cite{Mafra:2011kj} to construct BCJ numerators at $n$
points. All of this is applicable at tree level, whereas at loop level, the
relationship between open and closed string \emph{amplitudes} becomes obscure.

At the integrand level, five-point amplitudes were recently discussed in
\cite{Green:2013bza} in open and closed string. The authors of that work studied
how the closed-string integrand is related to the square of the open-string
integrand, and observed a detailed squaring behavior. They also discussed the
appearance of left-right mixing terms in this context. These terms are central 
in our one-loop analysis, even though at the qualitative level, four-points
amplitudes in $(\cN\!=\!2)\times(\cN\!=\!2)$ are more closely related to
six-point
ones in ${(\cN\!=\!4)\times(\cN\!=\!4)}$.

\section{Review of tree level in string theory}
\label{sec:st-tree}

In this section, we review RNS string amplitude calculations at tree level
in order to perform explicitly a five-point heterotic and type II computation,
as a warm-up exercise before going to the loop level.
Type I and II string amplitudes are known at $n$ points from the pure spinor formalism
\cite{Mafra:2010jq,Mafra:2010gj,Mafra:2011nv,Mafra:2011nw} and their field
theory limits were extensively studied in
\cite{Mafra:2011nw,Broedel:2013tta}, as well as their $\alpha'$ expansion in 
\cite{Schlotterer:2012ny,Broedel:2013aza,Broedel:2013tta,Stieberger:2013wea}.
As observed in \cite{BjerrumBohr:2010zs,Tye:2010dd}, the important point here is
not to focus
on the actual string theory amplitude, but rather to realize different
field theory limits by plugging different CFT's in the left- and right-moving
sectors of the string.
In that context, an observation that we shall make is that the
Mafra-Schlotterer-Stieberger open-string chiral block representation introduced in
\cite{Mafra:2011kj} to compute BCJ numerators can be used to construct directly
gravity amplitudes and make the double copy property manifest.
We perform this explicitly in the five-point case and briefly outline an
$n$-point extension.

Let us start from the integral for the five-particle scattering amplitude:
\begin{equation}
\cA^{\rm string}_5 = |z_{14}z_{45}z_{51}|^2
                     \int \d^2 z_2 \d^2 z_3
                     \langle V_1(z_1) V_2(z_2) V_3(z_3)
                             V_4(z_4) V_5 (z_5) \rangle \,,
\label{e:5ptsampcorr}
\end{equation}
where $|z_{14}z_{45}z_{51}|^2$ is the classical $c \bar c$ ghost correlator,
and we use the conformal gauge freedom to set $z_1=0,z_4=1,z_5\rightarrow\infty$.
The unintegrated vertex operators have a holomorphic and an anti-holomorphic part:
\begin{equation}
 V(z) = \, :V^{\rm (L)}(z) V^{\rm (R)}(\bar z) e^{ikX(z,\bar z)} : \,,
\end{equation}
where $V^{\rm (L)}$ and $V^{\rm (R)}$ are the chiral vertex operators for the left- and
right-moving sectors.\footnote{The chiral fields $X(z)$ and $X(\bar z)$ are
defined to contain half of the zero modes of the field
$X(z,\bar z) = x_0 + X_L(z) + X_R(\bar z))$ so that
$X(z) = x_0/2+X_L(z)$ and $X(\bar z) = x_0/2 +  X_R(\bar z)$.}
The notation for superscripts  $\rm (L)$ and $\rm (R)$ coincides with the one used
in \cite{Tye:2010dd}. Now, depending on what CFT we plug in these two
sectors, different theories in the low-energy limit can be realized, as
summarized in table~\ref{tab:models}. 
The anti-holomorphic vertex operators for the color CFT are gauge currents
\begin{equation}
   V^{\rm (R)}(\bar z) = T^a J_a(\bar z) \,,
\label{e:VOcurrent}
\end{equation}
where the $T^a$ matrices are in the adjoint representation
of the gauge group under consideration
(for instance, $E_8\times E_8$ or $SO(32)$ in the heterotic string
or more standard $SU(N)$ groups, after proper gauge group breaking by compactification).
The chiral vertex operators in the spacetime supersymmetric CFT
have a superghost picture number, $(-1)$ or $(0)$,
required to cancel the $(+2)$ background charge:
\begin{subequations} \begin{align}
  V^{\rm (L)}_{(-1)}(z) & = \varepsilon_{\mu}(k)\,  e^{-\phi}  \psi^\mu\,,
\label{e:VOsusy-1} \\
  V^{\rm (L)}_{(0)}(z)  & = \sqrt{\frac{2}{\ap}} \varepsilon_{\mu}(k)\,
      \Big(i \partial X^\mu + \frac \ap 2 (k \cdot \psi) \psi^\mu \Big)\,,
\label{e:VOsusy0}
\end{align} \label{e:VOsusy} \end{subequations}
where $\varepsilon_{\mu}(k)$ is the gluon polarization vector.
Therefore, at tree level, exactly two vertex operators must be chosen in the
$(-1)$ picture.

The anti-holomorphic vertex operators are then obtained from the holomorphic ones
by complex conjugation.
The total vertex operators of gluons and gravitons are constructed
as products of the chiral ones in accordance with table~\ref{tab:models}, and
the polarization tensor of the graviton is defined by the symmetric traceless
part of the product
$\varepsilon_{\mu\nu}(k)=\varepsilon_{\mu}(k)\varepsilon_{\nu}(k)$.

The correlation function \eqref{e:5ptsampcorr} can be also computed
as a product of a holomorphic and an anti-holomorphic correlator
thanks to the ``canceled propagator argument''. As explained in the classical
reference \cite[sec.~6.6]{Polchinski:1998rq}, the argument is essentially
an analytic continuation which makes sure that Wick contractions between
holomorphic and anti-holomorphic operators
\begin{equation}
	\langle \partial X(z,\bar z) \bar \partial X(w,\bar w) \rangle 
      = -\ap \pi \delta^{(2)}(z-w) \,,
\label{e:lr-tree-level}
\end{equation}
provide only vanishing contributions at tree level.\footnote{At one loop, a
zero mode term modify the right-hand side of eq.~\eqref{e:lr-tree-level}, see
eq.~\eqref{e:lr-one-loop}. This brings non-vanishing contributions,
whose analysis of the relationship with the BCJ construction is one of the aims
of this paper.}

Therefore, the chiral correlators can be dealt with separately. 
Our goal is to write them in the MSS chiral block representation~\cite{Mafra:2011kj},
in which
\begin{subequations} \begin{align}
\label{e:chirblockL}
\langle V^{\rm (L)}_1 V^{\rm (L)}_2 V^{\rm (L)}_3
        V^{\rm (L)}_4 V^{\rm (L)}_5 \rangle & = \!
\left(
\frac{ a^{\rm (L)}_1 }{z_{12} z_{23}}\!+\!
\frac{ a^{\rm (L)}_2 }{z_{13} z_{23}}\!+\!
\frac{ a^{\rm (L)}_3 }{z_{12} z_{34}}\!+\!
\frac{ a^{\rm (L)}_4 }{z_{13} z_{24}}\!+\!
\frac{ a^{\rm (L)}_5 }{z_{23} z_{34}}\!+\!
\frac{ a^{\rm (L)}_6 }{z_{23} z_{24}}
\right) \,, \\
\label{e:chirblockR} 
\langle V^{\rm (R)}_1 V^{\rm (R)}_2 V^{\rm (R)}_3
        V^{\rm (R)}_4 V^{\rm (R)}_5 \rangle & = \!  
\left(
\frac{ a^{\rm (R)}_1 }{\bar z_{12} \bar z_{23}}\!+\!
\frac{ a^{\rm (R)}_2 }{\bar z_{13} \bar z_{23}}\!+\!
\frac{ a^{\rm (R)}_3 }{\bar z_{12} \bar z_{34}}\!+\!
\frac{ a^{\rm (R)}_4 }{\bar z_{13} \bar z_{24}}\!+\!
\frac{ a^{\rm (R)}_5 }{\bar z_{23} \bar z_{34}}\!+\!
\frac{ a^{\rm (R)}_6 }{\bar z_{23} \bar z_{24}}
\right) \,, 
\end{align} \label{e:chirblocks} \end{subequations}
where $a^{\rm (L/R)}$ are independent of $z_i$ and carry either color or
kinematical information.
Accordingly, they are constructed either from the structure constants $f^{abc}$
of the gauge group 
or from momenta $k_i$ and polarization vectors $\varepsilon_i$ of the external states.
The plane-wave correlator (known as the Koba-Nielsen factor) writes
\begin{equation}
  \exp \Big(-\sum_{i<j} k_i \cdot k_j
             \langle X(z_i, \bar z_i) X(z_j, \bar z_j) \rangle \Big)
           = \prod_{i<j}|z_{ij}|^{\alpha' k_i \cdot k_j} \,,
\label{e:KNtree}
\end{equation}
where the bosonic correlator is normalized as follows:
\begin{equation}
      \langle X^\mu(z,\bar z) X^\nu(w,\bar w) \rangle
           = -{\ap\over 2} \eta^{\mu\nu} \ln(|z-w|^2) \,.
\label{e:Xcorr}
\end{equation}
It was implicitly taken into account when writing eqs.~\eqref{e:chirblocks},
since we included all possible Wick contractions, including those of the form
$\langle \partial X e^{ikX} \rangle$.

As we will see, taking the limit $\apt$ of eq.~\eqref{e:5ptsampcorr}
will lead us to the BCJ construction for the field theory amplitudes.
Note that if one derives $a^{\rm (L/R)}$ in a completely covariant way,
as is done in \cite{Mafra:2011kj}, one eventually obtains the BCJ numerators
valid in any dimension.
In this way, the whole BCJ construction can be regarded as a mere consequence
of the worldsheet structure in the low-energy limit.

In the following, we review the case of a correlator of anti-holomorphic gauge
currents, then we go to the supersymmetric kinematic sector.

\subsection{Gauge current correlators}
\label{sec:currcorr}

At level one, the current correlators of the Kac-Moody algebra of a given gauge
group are built from the standard OPE's:
\begin{equation}
 J^{a}(\bar z) J^b(0) = \frac{ \delta^{ab}}{\bar z^2}
                      + i f^{a b c}\frac{J^c(\bar z)}{\bar z} + ... \,,
\end{equation}
where the $f^{abc}$'s are the structure constants of the gauge group,
defined by 
\begin{equation}
      [T^a,T^b]=i f^{abc} T^c \,.
\label{e:fabc}
\end{equation}

At four points, one can thus obtain the following correlator:
\begin{equation}
  \langle J^{a_1}(\bar z_1)J^{a_2}(\bar z_2)
          J^{a_3}(\bar z_3)J^{a_4}(\bar z_4)\rangle
= \frac{ \delta^{a_1 a_2}\delta^{a_3 a_4}}{\bar z_{12}^2 \bar z_{34}^2} -
  \frac{f^{a_1 a_2b} f^{ba_3a_4}}
  {\bar z_{12}\bar z_{23}\bar z_{34}\bar z_{41}}
+ (2\leftrightarrow3) + (2\leftrightarrow4) \,,
\label{e:currcorr}
\end{equation}
where the conformal gauge is not fixed.
In the low-energy limit of the heterotic string amplitude,
the $\delta \delta$-terms in \eqref{e:currcorr} produce
the non-planar contribution of the gravity sector (singlet exchange),
while the $ff$-terms result in the gluon exchange channel.
In the following, we shall decouple these non-planar corrections by hand.

At five points,
to obtain the correct MSS chiral blocks for the gauge current correlator,
one only needs to repeatedly use the Jacobi identities \eqref{e:jacobi0}.
After fixing the conformal gauge by setting $z_1=0,z_4=1,z_5\rightarrow\infty$, we get
\begin{equation} \begin{aligned}
   \langle J^{a_1}(\bar z_1) J^{a_2}(\bar z_2)
           J^{a_3}(\bar z_3) J^{a_4}(\bar z_4)
           J^{a_5}(\bar z_5) \rangle = \,&
{f^{a_1a_2b}f^{ba_3c}f^{ca_4a_5}\over \bar z_{12}\bar z_{23}} -
{f^{a_1a_3b}f^{ba_2c}f^{ca_4a_5}\over \bar z_{13}\bar z_{23}} \\ -
{f^{a_1a_2b}f^{ba_5c}f^{ca_3a_4}\over \bar z_{12}\bar z_{34}} -
{f^{a_1a_3b}f^{ba_5c}f^{ca_2a_4}\over \bar z_{13}\bar z_{24}} + \,&
{f^{a_1a_5b}f^{ba_2c}f^{ca_4a_3}\over \bar z_{23}\bar z_{34}} -
{f^{a_1a_5b}f^{ba_3c}f^{ca_4a_2}\over \bar z_{23}\bar z_{24}} \\ + \,&
\text{non-planar terms} \,, \end{aligned}
\label{e:5currents}
\end{equation}
Now we can immediately read off the following set of 6 color factors:
\begin{equation} \begin{aligned}
      a_1^{\rm (R)}&=f^{a_1a_2b}f^{ba_3c}f^{ca_4a_5}\,,\quad 
      a_2^{\rm (R)}&=f^{a_1a_3b}f^{ba_2c}f^{ca_4a_5}\,,\quad 
      a_3^{\rm (R)}&=f^{a_1a_2b}f^{ba_5c}f^{ca_3a_4}\,,\\
      a_4^{\rm (R)}&=f^{a_1a_3b}f^{ba_5c}f^{ca_2a_4}\,,\quad
      a_5^{\rm (R)}&=f^{a_1a_5b}f^{ba_2c}f^{ca_4a_3}\,,\quad
      a_6^{\rm (R)}&=f^{a_1a_5b}f^{ba_3c}f^{ca_4a_2}\,.
\end{aligned} \label{e:5fff} \end{equation}
It actually corresponds to the color decomposition into $(n-2)!$ terms
uncovered in \cite{DelDuca:1999rs}.

\subsection{Kinematic CFT}
\label{sec:kincorr}

Now let us compute the RNS 5-point left-moving correlator in the
supersymmetric sector,
\begin{equation}
 \langle V^{\rm (L)}_{(-1)}(z_1) V^{\rm (L)}_{(0)}(z_2) V^{\rm (L)}_{(0)}(z_3)
         V^{\rm (L)}_{(0)}(z_4) V^{\rm (L)}_{(-1)}(z_5) \rangle \,,
\label{e:Lcorr}
\end{equation}
where the chiral vertex operators for the kinematic CFT were defined in \eqref{e:VOsusy}.
In \eqref{e:Lcorr}, we picked two vertex operators to carry ghost picture number $(-1)$
in such a way that all double poles can be simply eliminated by a suitable gauge choice.
The correlator \eqref{e:Lcorr} is computed using Wick's theorem
along with the two-point function \eqref{e:Xcorr} and
\begin{subequations} \begin{align}
   \langle \psi^\mu(z)\psi^\nu(w) \rangle
      & = \eta^{\mu\nu}/(z-w) \,,
\label{e:psicorr} \\
   \langle \phi(z) \phi(w)\rangle & = -\ln(z-w) \,.
\label{e:phicorr}
\end{align} \label{e:correlators} \end{subequations}
For a completely covariant calculation, we refer the reader to \cite{Mafra:2011kj},
whereas here for simplicity we restrict ourselves to
the MHV amplitude $\cA(1^+,2^-,3^-,4^+,5^+)$
with the following choice of reference momenta:
\begin{equation}
   (q^{\rm ref}_1,q^{\rm ref}_2,q^{\rm ref}_3,q^{\rm ref}_4,q^{\rm ref}_5)
      = (k_2,k_1,k_1,k_2,k_2) \,.
\label{e:refmomenta}
\end{equation}
In combination with the ghost picture number choice,
this gauge choice eliminates a lot of terms and, in particular, all double poles.
We end up with only ten terms
of the form\footnote{See the full expression in appendix~\ref{app:numerators}.}
\begin{equation*}
  \frac{(\varepsilon_3 \varepsilon_5) (\varepsilon_1 k_3)
        (\varepsilon_2 k_3) (\varepsilon_4 k_1)}
       {8 z_{1 3} z_{2 3}}
- \frac{(\varepsilon_3 \varepsilon_5) (\varepsilon_1 k_3)
        (\varepsilon_2 k_4) (\varepsilon_4 k_3)}
       {8 z_{1 3} z_{2 4} z_{3 4}} + \dots \,.
\end{equation*}

To reduce them to the six terms of the MSS chiral block representation,
one could apply in this closed-string context the open string technology based
on repeated worldsheet IBP's described in
\cite{Mafra:2011nv,Mafra:2011nw,Broedel:2013aza,Broedel:2013tta}.
However, the situation is greatly simplified here,
since we have already eliminated all double poles.
Thanks to that, we can proceed in a pedestrian way
and only make use of partial fractions identities, such as
\begin{equation}
   \frac{1}{z_{12}z_{24}z_{34}} =-\frac{z_{12}+z_{24}}{z_{12}z_{24}z_{34}} \,,
\label{e:part-frac}
\end{equation}
where we take into account that $z_{41}=1$.
Our final result, similarly to the one in appendix~D of \cite{Mafra:2011kj},
contains two vanishing and four non-vanishing coefficients.
In the spinor-helicity formalism, they are
\begin{equation} \begin{aligned}
  a_1^{\rm (L)} = 0 & \,,\quad &
  a_2^{\rm (L)} = & \,\,
      {\spaa23^4 \spbb31^2 \spbb54 \over \spaa24 \spaa25 \spaa12  \spbb21}, \quad &
  a_3^{\rm (L)}=0 & \,, \\
  a_4^{\rm (L)} = \phantom{0} & \!\!\!
      {\spaa 23^3 \spbb31 \spbb41 \spbb54 \over \spaa12 \spaa25 \spbb21} \,, &
  a_5^{\rm (L)}= &
      {\spaa23^3 \spbb51^2 \spbb43 \over \spaa24 \spaa12 \spbb21} \,, \quad &
  a_6^{\rm (L)}= \phantom{0} & \!\!\!
      {\spaa23^3 \spbb41 \spbb43 \spbb51 \over \spaa12 \spaa25 \spbb21} \,.
\end{aligned} \label{e:kinblocks} \end{equation}

\subsection{Low-energy limit}
\label{sec:low-energy-limit}

\begin{figure}[t]
 \centering\includegraphics[scale=0.75]{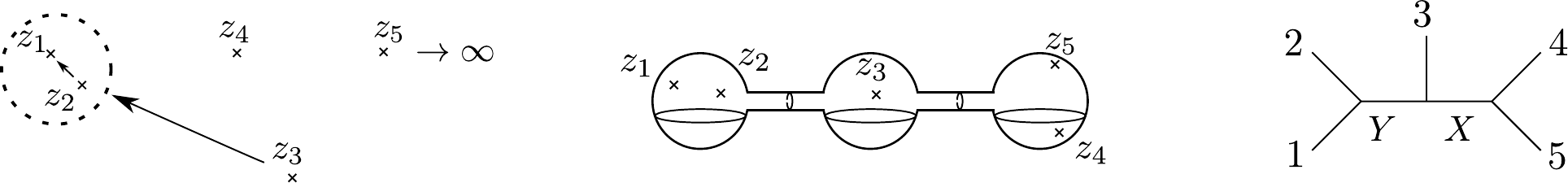}
 \caption{Five-point low-energy limit: the complex plane,
the tubed worldsheet and the tropical (worldline) graph.
$X$ and $Y$ are the lengths of the edges, or proper-time variables.}
 \label{fig:5ptsdegen}
\end{figure}

Before specializing to a particular theory (gauge theory or gravity),
let us review the general low-energy limit mechanism at tree level. 
In the open string, very efficient procedures have been developed for extracting
the low-energy limit of $n$-points amplitudes in a systematic way
\cite{Mafra:2011nw,Broedel:2013tta}. 
The essential point, common to both open and closed string procedures,
consists in the observation that a pole\footnote{We define Mandelstam kinematic
invariants $s_{ij}$ in the $(+,-,-,-)$ signature by $s_{ij}=(k_i+k_j)^2$.} in
the channel $s_{ij} s_{kl}$ comes from integrating over the region of the moduli
space where $z_i$ and $z_k$ collide to $z_j$ and $z_l$, respectively,
provided that the integrand contains a pole in the variables $z_{ij} z_{kl}$.
In these regions, the closed string worldsheet looks like spheres
connected by very long tubes (see figure~\ref{fig:5ptsdegen}), and we simply have to
integrate out the angular coordinates along the tubes to obtain graph edges.
This is the basic mechanism of the tropical limiting procedure reviewed in
\cite{Tourkine:2013rda} (see section VI.A for four-tachyon and four-graviton
examples).

A slight subtlety to take into account is that the $s_{45}$-channel pole, for
instance, is not due to the pole $1/z_{45}$, as both $z_4$ and $z_5$ are fixed
and $z_5=\infty$ is already absent from the expressions. 
Rather, it is created when both $z_2, z_3 \rightarrow z_1$,
i.e. it appears as a $s_{123}$ pole. 
Moreover, the details of the double limit matter as well: suppose we want the
kinematic pole $1/(s_{12}s_{123})$, it is generated by the \emph{successive}
limit $z_2 \rightarrow z_1$ and then $z_3$ to the cluster formed by $z_1$ and $z_2$.
In other words, it arises from regions of the moduli space where
$|z_{12}| \ll |z_{23}| \sim |z_{13}| \ll 1$.
We can describe the geometry of the worldsheet in this limit
by going to the following tropical variables:
\begin{equation}
 z_{13} = -e^{i \theta} e^{-X/\ap}, \qquad
 z_{12} = z_{13}e^{i \phi} e^{-Y/\ap}= -e^{i(\theta+\phi)}e^{-(X+Y)/\ap} ,
\label{e:tropch}
\end{equation}
where $X$ and $Y$ are the lengths (or the Schwinger proper time variables)
of the edges of the graph as pictured in figure~\ref{fig:5ptsdegen}.
The phases $\theta$ and $\phi$ are the cylindrical coordinates along the tubes
that need to be integrated out in order to recover
a purely one-dimensional variety corresponding to a Feynman graph.

Accordingly, the integration measure produces a Jacobian:
\begin{equation}
 \d^2 z_2  \to -{1\over \ap } \d X \d \theta \, e^{-2X/\ap} , \quad
 \d^2 z_3  \to -{1\over \ap } \d Y \d \phi \, e^{-2(X+Y)/\ap} .
\label{e:jacobian}
\end{equation}
Then one can check that the exponential factor transforms as follows:
\begin{equation}
   e^{ \ap (k_1 \cdot k_2) \ln|z_{12}| + \ap (k_1 \cdot k_3) \ln|z_{13}|
                                       + \ap (k_2 \cdot k_3) \ln|z_{23}| }
 = e^{-X s_{123}/2 - Y s_{12}/2 } +O(\ap) ,
\end{equation}
where the phase dependence in $\theta$ and $\psi$ is trivial, granted that $Y$
and $X$ are greater than some UV cutoff of order $\ap$. At this point, we have
integrands of the form
\begin{equation}
\frac{1}{\ap^2} \d X \d Y \d \theta \d \phi
\frac{e^{-2X/\ap}e^{-2(X+Y)/\ap}}{z_{ij}z_{kl} \bar z_{mn} \bar z_{pq}} e^{-X
s_{123}/2 - Y s_{12}/2 }\,.
\end{equation}
To make them integrate to the expected double pole $1/s_{12}s_{123}$,
we need the Jacobian to be compensated by the $z_{ij}$ poles in such a way
that we can integrate out the phases $\theta$ and $\phi$. 
This is carried out by writing the amplitude \eqref{e:5ptsampcorr}
in the MSS chiral block representation:
\begin{equation} \begin{aligned}
{\cal A}_5^{{\rm string}}(1^+,2^-,3^-,4^+,5^+;\alpha') =
      (2\ap)^{2} \int \! \!  {\rm d} z_2 \, {\rm d} z_3 \, \prod_{i<j}
      |z_{ij}|^{ \alpha' k_i \cdot k_j} & \\
      \times \left( \frac{ a^{\rm (L)}_1 }{z_{12} z_{23}} 
                  + \frac{ a^{\rm (L)}_2 }{z_{13} z_{23}}
                  + \frac{ a^{\rm (L)}_3 }{z_{12} z_{34}}
                  + \frac{ a^{\rm (L)}_4 }{z_{13} z_{24}}
                  + \frac{ a^{\rm (L)}_5 }{z_{23} z_{34}}
                  + \frac{ a^{\rm (L)}_6 }{z_{23} z_{24}} \right) & \\
      \times \left( {a^{\rm (R)}_1 \over \bar z_{12}\bar z_{23}}
                  + {a^{\rm (R)}_2 \over \bar z_{13}\bar z_{23}}
                  + {a^{\rm (R)}_3 \over \bar z_{12}\bar z_{34}}
                  + {a^{\rm (R)}_4 \over \bar z_{13}\bar z_{24}}
                  + {a^{\rm (R)}_5 \over \bar z_{23}\bar z_{34}}
                  + {a^{\rm (R)}_6 \over \bar z_{23}\bar z_{24}} \right) & .
\label{e:5ptcorr}
\end{aligned} \end{equation}
It is not difficult to convince oneself that the only terms that do
not vanish in this particular limit, where $|z_{12}|\ll|z_{23}|\ll1$, are
exactly the products of $1/|z_{12}|^2$ with any of the following:
$1/|z_{23}|^2$, $1/(z_{23} \bar z_{13})$, $1/(z_{13} \bar z_{23})$, or $1/|z_{13}|^2$,
since $1/z_{13} = 1/z_{23} + O(e^{-Y/\ap})$.
Any of these terms obviously cancel the Jacobian.
Moreover, they do not vanish when the phase integration is performed.
If instead one had a term like $1/(z_{12} \bar z_{12}^3)=e^{2X/\ap}e^{i \theta}$,
it would cancel $e^{-2X/\ap}$ in the Jacobian
but would vanish after integration over $\theta$.

It is a characteristic feature of the MSS representation that only the terms with the
correct weight are non-zero upon phase integration. That is why it
is particularly suitable for the analysis of the low-energy limit of the closed
string. In other words, the phase dependence is trivial by construction, which
means that the level-matching is automatically satisfied.
To sum up, to obtain a pole in $1/s_{12}s_{123}$, we have to pick up exactly two
chiral blocks $1/z_{12}z_{23}$ and $1/\bar z_{12}\bar z_{23}$ in
\eqref{e:5ptcorr} which come with a factor of $a_1^{(L)} a_1^{(R)}$.
Furthermore, it can be easily proven that any other region of the moduli space,
where at least one of $z_2$ or $z_3$ stay at finite distance from other points,
contributes only to subleading $O(\ap)$ terms.
In total, this region of the moduli space contributes the following pole
to the amplitude:
\begin{equation}
      \frac{ a_1^{\rm (L)} a_1^{\rm (R)}}{ s_{12}s_{123} } \,,
\end{equation}
where one reads $n_1^{\rm (L/R)} = a_1^{\rm (L/R)}$.
One can then repeat this operation in the other kinematic channels.
For instance,\footnote{The channels generated by $z_2$ and/or $z_3\to
z_5=\infty$
are dealt with by introducing a $+$ sign in the exponential in \eqref{e:tropch}.
Then the pole is generated by a similar procedure}
the region $z_{23}\ll z_{34}\sim z_{24}\ll1$ receives non-zero contributions both
from $1/(z_{23} z_{34})$ and $1/(z_{23} z_{24})$ (and their complex conjugates).
This results in the following contribution to the low-energy limit of the amplitude:
\begin{equation}
      \frac{(a_5^{\rm (L)}+a_6^{\rm (L)})(a_5^{\rm (R)}+a_6^{\rm (R)})}
           {s_{23}s_{234}} \,.
\end{equation}

By repeating this operation in all other kinematic channels, 
one can generate the $15$ combinatorially-distinct trivalent graphs
of the low-energy limit and thus obtain the full field theory amplitude,
valid in any dimension:
\begin{equation} \begin{aligned}
{\cal A}^{\rm tree}_5 =  g^3
 \biggl( & {n^{\rm (L)}_1 n^{\rm (R)}_1 \over s_{12}s_{45}}
        + {n^{\rm (L)}_2 n^{\rm (R)}_2 \over s_{23}s_{51}}
        + {n^{\rm (L)}_3 n^{\rm (R)}_3 \over s_{34}s_{12}}
        + {n^{\rm (L)}_4 n^{\rm (R)}_4 \over s_{45}s_{23}}
        + {n^{\rm (L)}_5 n^{\rm (R)}_5 \over s_{51}s_{34}}
        + {n^{\rm (L)}_6 n^{\rm (R)}_6 \over s_{14}s_{25}} \\
      + & {n^{\rm (L)}_7 n^{\rm (R)}_7 \over s_{32}s_{14}}
        + {n^{\rm (L)}_8 n^{\rm (R)}_8 \over s_{25}s_{43}}
        + {n^{\rm (L)}_9 n^{\rm (R)}_9 \over s_{13}s_{25}}
        + {n^{\rm (L)}_{10} n^{\rm (R)}_{10} \over s_{42}s_{13}}
        + {n^{\rm (L)}_{11} n^{\rm (R)}_{11} \over s_{51}s_{42}}
        + {n^{\rm (L)}_{12} n^{\rm (R)}_{12} \over s_{12}s_{35}} \\
      + & {n^{\rm (L)}_{13} n^{\rm (R)}_{13} \over s_{35}s_{24}}
        + {n^{\rm (L)}_{14} n^{\rm (R)}_{14} \over s_{14}s_{35}}
        + {n^{\rm (L)}_{15} n^{\rm (R)}_{15} \over s_{13}s_{45}} \biggr)\,,
\label{e:5pttree}
\end{aligned} \end{equation}
in terms of the following numerators:
\begin{equation} \begin{aligned}
& n^{\rm (L/R)}_{1 }\!=   a^{\rm (L/R)}_1 \,,
~~~~~~~~~~~~~~~~~~~~~~~~~~~~~~~~~~~~~~~~~~~~~\;\,
  n^{\rm (L/R)}_{2 }\!=   a^{\rm (L/R)}_5\!+ a^{\rm (L/R)}_6 \,, \\
& n^{\rm (L/R)}_{3 }\!= - a^{\rm (L/R)}_3 \,,
~~~~~~~~~~~~~~~~~~~~~~~~~~~~~~~~~~~~~~~~~~~~
  n^{\rm (L/R)}_{4 }\!=   a^{\rm (L/R)}_1\!+ a^{\rm (L/R)}_2 \,, \\
& n^{\rm (L/R)}_{5 }\!=   a^{\rm (L/R)}_5 \,,
~~~~~~~~~~~~~~~~~~~~~~
  n^{\rm (L/R)}_{6 }\!= - a^{\rm (L/R)}_2\!- a^{\rm (L/R)}_4
                      \!+ a^{\rm (L/R)}_3\!+ a^{\rm (L/R)}_5 \,, \\
& n^{\rm (L/R)}_{7 }\!= - a^{\rm (L/R)}_1\!- a^{\rm (L/R)}_2
                      \!+ a^{\rm (L/R)}_5\!+ a^{\rm (L/R)}_6 \,, ~~~~~~~~~~~
  n^{\rm (L/R)}_{8 }\!=   a^{\rm (L/R)}_3\!+ a^{\rm (L/R)}_5 \,, \\
& n^{\rm (L/R)}_{9 }\!= - a^{\rm (L/R)}_2\!- a^{\rm (L/R)}_4 \,,
~~~~~~~~~~~~~~~~~~~~~~~~~~~~~~~~\:\,
  n^{\rm (L/R)}_{10}\!= - a^{\rm (L/R)}_4 \,, \\
& n^{\rm (L/R)}_{11}\!=   a^{\rm (L/R)}_6 \,,
~~~~~~~~~~~~~~~~~~~~~~~~~~~~~~~~~~~~~~~~~~~~~~~\!
  n^{\rm (L/R)}_{12}\!=   a^{\rm (L/R)}_1\!+ a^{\rm (L/R)}_3 \,, \\
& n^{\rm (L/R)}_{13}\!=   a^{\rm (L/R)}_4\!+ a^{\rm (L/R)}_6 \,, ~~~~~~~~~~~
  n^{\rm (L/R)}_{14}\!= - a^{\rm (L/R)}_4\!- a^{\rm (L/R)}_6
                      \!+ a^{\rm (L/R)}_1\!+ a^{\rm (L/R)}_3 \,, \\
& n^{\rm (L/R)}_{15}\!= - a^{\rm (L/R)}_2 \,.
\label{e:BCJnums}
\end{aligned} \end{equation}

It is now trivial to check that, by construction, $n_i$'s satisfy the Jacobi identities,
which we recall in appendix~\ref{app:numerators}.
The linear relations \eqref{e:BCJnums} between $n_i$'s and $a_i$'s coincide with
those derived for gauge theory amplitudes in \cite{Mafra:2011kj},
where covariant expressions for the kinematical numerators at any multiplicity were obtained.

The crucial point here is that we have not referred to the actual expressions
of the $n_i$'s derived in the previous sections
but simply started from the MSS representation \eqref{e:5ptcorr} for the string amplitude.
Therefore, the final result \eqref{e:5pttree} can be either a gauge theory or a
gravity amplitude, depending on the type of string theory in which the
low-energy limit is taken,
as indicated in table \ref{tab:models}.
In heterotic string theory,
if $n_i^{\rm (L)}=c_i$ are color factors and $n_i^{\rm (R)}=n_i$ are kinematic numerators,
one obtains a gluon amplitude,
whereas in type II string theory,
where both $n_i^{\rm (L)}$ and $n_i^{\rm (R)}$ are kinematic numerators,
one gets a graviton scattering amplitude.

Another option would be to choose both $n_i^{\rm (L)}$ and $n_i^{\rm (R)}$ to be
color factors $c_i$, in which case \eqref{e:5pttree} would correspond
to the scattering amplitude of five color cubic scalars.
From the perspective of the low-energy limit of string theory, this corresponds
to compactifying both sectors of the bosonic string
on the same torus as the one of the heterosis mechanism
and then choosing external states bipolarized in the internal color space. 
This string theory, of course, suffers from all known inconsistencies
typical for the bosonic string.
However, at tree level, if one decouples by hand in both sectors the terms
which create non-planar corrections in the heterotic string,
the pathological terms disappear.

Therefore, the formula \eqref{e:5pttree} can be extended to produce
the tree-level five-point amplitudes of the three theories: gravity, Yang-Mills and
cubic scalar color. This is done by simply choosing different target-space
polarizations for $\rm (L)$ and $\rm (R)$, as in table~\ref{tab:models},
to which, in view of the previous discussion,
we could now add a new line for the cubic scalar color model.

The point of this demonstration was to illustrate the fact that
the product of the left- and right-moving sectors produces in the low-energy limit
the form of the amplitude in which the double copy construction is transparent
and is not a peculiarity of gravity but rather of any of the three theories.
This suggests that both the BCJ duality in gauge theory
and the double copy construction of gravity follow from the inner structure
of the closed string and its low-energy limit.

Furthermore, the MSS chiral block representation exists
for $n$-point open string amplitudes
\cite{Mafra:2011kj,Mafra:2011nv,Mafra:2011nw},
so to extend those considerations to any multiplicity, one would only need
to rigorously prove that any open string pole channel corresponds to a closed string one
and verify that level matching correctly ties the two sectors together.
Then the MSS construction would produce the BCJ construction at any multiplicity,
and this would constitute a string-theoretic proof that the BCJ representation
of Yang-Mills amplitudes implies the double copy construction of gravity amplitudes
at tree level.
Finally, note that this procedure is different from the KLT approach \cite{Kawai:1985xq}
in that it relates the numerators of cubic diagrams in the various theories,
rather than the amplitudes themselves.
All of this motivates our study of the double copy construction at higher loops
in the purely closed string sector.

We conclude this section by the observation that, in the recent works related to the
``scattering equations'' \cite{Cachazo:2013iea,Cachazo:2013gna,Cachazo:2013hca,Monteiro:2013rya,
Mason:2013sva,Berkovits:2013xba,Dolan:2013isa},
there appeared new formulas for tree-level scattering amplitudes
of cubic color scalars, gauge bosons and gravitons,
in which color and kinematics play symmetric roles.
It was also suggested that this approach might be generalizable to higher-spin amplitudes. 
Naturally, it would be interesting to make a direct connection between
the scattering equations and the approach based on the low-energy limit of the closed string.

\section{One loop in field theory}
\label{sec:ft-loop}

      In this section, we turn to the study of the BCJ duality at one loop.
Here and in the rest of this paper, we will deal only with amplitudes
with the minimal number of physical external particles in supersymmetric
theories -- four.\footnote{Higher multiplicity amplitudes in maximal SYM and
supergravity have been addressed in the context of the BCJ duality in the
upcoming paper \cite{Mafra:2014oia,He:XX}.}
At one loop, a color-dressed four-gluon amplitude can be represented as
\begin{equation} \begin{aligned}
      \mathcal{A}^{\text{1-loop}}(1^-,2^-,3^+,4^+)
      = \!\!\int\! \frac{\d^d \ell}{(2\pi)^d}
        \bigg\{
            & \, f^{b a_1 c} f^{c a_2 d} f^{d a_3 e} f^{e a_4 b}
             \frac{n^{\text{box}}(1^-,2^-,3^+,4^+; \ell)}
                  {\ell^2 (\ell-k_1)^2 (\ell-k_1-k_2)^2 (\ell+k_4)^2 } \\
          + & \, f^{b a_1 c} f^{c a_2 d} f^{d a_4 e} f^{e a_3 b}
             \frac{n^{\text{box}}(1^-,2^-,4^+,3^+; \ell)}
                  {\ell^2 (\ell-k_1)^2 (\ell-k_1-k_2)^2 (\ell+k_3)^2 } \\
          + & \, f^{b a_1 c} f^{c a_4 d} f^{d a_2 e} f^{e a_3 b}
             \frac{n^{\text{box}}(1^-,4^+,2^-,3^+; \ell)}
                  {\ell^2 (\ell-k_1)^2 (\ell-k_1-k_4)^2 (\ell+k_3)^2 } \\
          + & \, \text{triangles, etc.}
        \bigg\} \,.
\label{colorstructure}
\end{aligned} \end{equation}
Recall that the color factors can also be written in terms of color traces,
for example:      
\begin{equation} \begin{aligned}
      c^{\text{box}}(1^-,2^-,3^+,4^+)
       & \equiv f^{b a_1 c} f^{c a_2 d} f^{d a_3 e} f^{e a_4 b} \\
       & = N_c \left( \tr(T^{a_1} T^{a_2} T^{a_3} T^{a_4})
                    + \tr(T^{a_4} T^{a_3} T^{a_2} T^{a_1}) \right)
              + \text{double traces} \,.
\label{colorfactor}
\end{aligned} \end{equation}
In this way, one can easily relate the color-kinematics representation \eqref{e:Ageneral}
to the primitive amplitudes that are defined as the coefficients of the leading
color traces \cite{Bern:1990ux}.

\subsection{Double copies of one $\cN=4$ SYM}
\label{sec:ft-N4}

The maximally supersymmetric Yang-Mills theory has the simplest BCJ numerators.
At four points, they are known up to four loops
\cite{Bern:1998ug,Bern:2010ue,Bern:2012uf}, and only at three loops they start
to depend on loop momenta, in accordance with the string theory understanding
\cite{Green:2006gt,Berkovits:2009aw,Gomez:2013sla}.
For example, the one-loop amplitude is just a sum of three scalar boxes
\cite{Green:1982sw},
which is consistent with the color-kinematic duality in the following way:
the three master boxes written in \eqref{colorstructure}
have the same trivial numerator
$\braket{12}^2 [34]^2 =  i st A^{tree}(1^-,2^-,3^+,4^+)$
(which we will always factorize henceforward),
and all triangle numerators are equal to zero by the Jacobi identities.

Thanks to that particularly simple BCJ structure of $\cN=4$ SYM,
the double copy construction for $\cN \geq 4$ supergravity amplitudes
simplifies greatly \cite{Bern:2011rj}.
Indeed, as the second gauge theory copy does not have to obey the BCJ duality,
one can define its box numerators simply by taking its entire planar integrands
and putting them in a sum over a common box denominator.
Since the four-point $\cN\!=\!4$ numerators are independent of the loop momentum,
the integration acts solely on the integrands of the second Yang-Mills copy
and thus produces its full primitive amplitudes:
      \begin{equation} \begin{aligned}
            \mathcal{M}_{\cN=\,4+N,\text{grav}}^{\text{1-loop}}(1^-,2^-,3^+,4^+) =
                  \braket{12}^2 [34]^2
                  \bigg\{ A_{\cN=N,\text{vect}}^{\text{1-loop}}&(1^-,2^-,3^+,4^+) \\
                     + \, A_{\cN=N,\text{vect}}^{\text{1-loop}} (1^-,2^-,4^+,3^+)
                        + A_{\cN=N,\text{vect}}^{\text{1-loop}}&(1^-,4^+,2^-,3^+)
                  \bigg\} \,.
      \label{N4plusgravity}
	\end{aligned} \end{equation}
The $\cN=8$ gravity amplitude is then simply given by the famous result of
\cite{Green:1982sw} in terms of the scalar box integrals $I_4$,
recalled in appendix~\ref{app:integrals}:
      \begin{equation} \begin{aligned}
            \mathcal{M}_{\cN=8,\text{grav}}^{\text{1-loop}}(1^-,2^-,3^+,4^+) =
                   \frac{i}{ (4\pi)^{d/2} } \braket{12}^4 [34]^4
                  \bigg\{ I_4(s,t) + I_4(s,u) + I_4(t,u) \bigg\} \,.
      \label{N8gravity}
	\end{aligned} \end{equation}

For a less trivial example, let us consider the case of the $\cN=6$ gravity,
for which the second copy is the contribution of a four-gluon scattering
in $\cN=2$ SYM.
It is helpful to use the one-loop representation of the latter as
\begin{equation}
      A^{\text{1-loop}}_{\cN=2,\text{vect}} =
      A^{\text{1-loop}}_{\cN=4,\text{vect}} - 2 \,
      A^{\text{1-loop}}_{\cN=2,\text{hyper}} \,,
\label{amplitudeN2}
\end{equation}
where the last term is the gluon amplitude contribution
from the $\cN=2$ hyper-multiplet
(or, equivalently, $\cN=1$ chiral-multiplet in the adjoint representation)
in the loop.
This multiplet is composed of two scalars and one Majorana spinor,
so its helicity content can be summarized as
$\left( 1_{\frac{1}{2}}, 2_{0}, 1_{-\frac{1}{2}} \right)$.
If we use eq.~\eqref{N4plusgravity}
to ``multiply'' eq.~\eqref{amplitudeN2} by $\cN=4$ SYM,
we obtain a similar expansion for the gravity amplitudes:
      \begin{equation}
            \mathcal{M}_{\cN=6,\text{grav}}^{\text{1-loop}} =
            \mathcal{M}_{\cN=8,\text{grav}}^{\text{1-loop}}
       - 2\,\mathcal{M}_{\cN=6,\text{matt}}^{\text{1-loop}} \,,
      \label{N6expansion}
	\end{equation}
where ``$\cN\!=\!6$ matter'' corresponds
to the formal multiplet which contains a spin-$3/2$ Majorana particle
and can be represented as
$\left( 1_{\frac{3}{2}}, 6_{1}, 15_{\frac{1}{2}}, 20_{0},
      15_{-\frac{1}{2}}, 6_{-1}, 1_{-\frac{3}{2}} \right)$.
Its contribution to the amplitude can be constructed through eq.~\eqref{N4plusgravity}
as $(\cN\!=\!4)\times(\cN\!=\!2\text{ hyper})$,
where the second copy is also well known \cite{Bern:1994cg,Bern:1995db}
and is most easily expressed in terms of scalar integrals $I_n$:
\begin{subequations} \begin{align}
      A_{\cN=2,\text{hyper}}^{\text{1-loop}}(1^-,2^-,3^+,4^+) =
            \frac{i}{ (4\pi)^{d/2} } \braket{12}^2 [34]^2
            \bigg\{& -\frac{1}{st} I_2(t) \bigg\}
\label{N2hyperSH} \,, \\
      A_{\cN=2,\text{hyper}}^{\text{1-loop}}(1^-,4^+,2^-,3^+) =
            \frac{i}{ (4\pi)^{d/2} } \braket{12}^2 [34]^2
            \bigg\{&  \frac{t u}{2s^2} I_4(t,u) \nonumber \\
                   +\!\frac{t}{s^2} I_3(t) + \frac{u}{s^2} I_3(u)
                   & +\frac{1}{st} I_2(t) + \frac{1}{su} I_2(u)
            \bigg\}
\label{N2hyperNSH} \,.
\end{align} \label{N1chiral} \end{subequations}
This lets us immediately write down the result from \cite{Bern:2011rj}:
	      \begin{equation}
            \mathcal{M}_{\cN=6,\text{matt}}^{\text{1-loop}}(1^-,2^-,3^+,4^+)
                  = \frac{i}{ (4\pi)^{d/2} } \braket{12}^4 [34]^4
                  \bigg\{
                        \frac{tu}{2s^2} I_4(t,u)
                      + \frac{t}{s^2} I_3(t) + \frac{u}{s^2} I_3(u)
                  \bigg\} \,.
      \label{N6gravitymatter}
	\end{equation}
A comment is due here: here and below, we use the scalar integrals $I_n$
recalled in appendix~\ref{app:integrals}, just as a way of writing down
integrated expressions, so the scalar triangles in eq.~\eqref{N6gravitymatter}
do not contradict with the no-triangle property of $\cN=4$ SYM.
As explained earlier, the BCJ double copy construction behind eq.~\eqref{N4plusgravity},
and its special case \eqref{N6gravitymatter}, contains only the box topology
with all the scalar integrals in eq.~\eqref{N1chiral} collected into non-scalar boxes.

Having thus computed
$ \mathcal{M}_{\cN=6}^{\text{1-loop}}(1^-,2^-,3^+,4^+) $
through the expansion~\eqref{N6expansion}, we can consider computing
      \begin{equation}
            \mathcal{M}_{\cN=4,\text{grav}}^{\text{1-loop}} =
            \mathcal{M}_{\cN=8,\text{grav}}^{\text{1-loop}}
       - 4\,\mathcal{M}_{\cN=6,\text{matt}}^{\text{1-loop}}
       + 2\,\mathcal{M}_{\cN=4,\text{matt}}^{\text{1-loop}} \,,
      \label{N4expansion}
	\end{equation}
where the $\cN=4$ matter multiplet
$\left( 1_{1}, 4_{\frac{1}{2}}, 6_{0}, 4_{-\frac{1}{2}}, 1_{-1} \right)$
can be constructed either through eq.~\eqref{N4plusgravity}
as $(\cN\!=\!4)\times(\cN\!=\!0\text{ scalar})$,
or as $(\cN\!=\!2\text{ hyper})^2$.
In the former case, one only needs the full amplitudes from \cite{Bern:1995db}
to obtain the following result \cite{Dunbar:1994bn,Bern:2011rj}:
      \begin{equation} \begin{aligned}
            \mathcal{M}_{\cN=4,\text{matt}}^{\text{1-loop}}(1^-,2^-,3^+,4^+)
                  = \frac{-i}{ (4\pi)^{d/2} } \braket{12}^4 [34]^4 
                  \bigg\{
                        \frac{tu}{s^3} I_4^{d=6-2\epsilon}(t,u)
                      - \frac{u}{s^3} I_2(t) - \frac{t}{s^3} I_2(u) & \\
                      - \frac{1}{s^2}
                        \left( I_3^{d=6-2\epsilon}(t)
                             + I_3^{d=6-2\epsilon}(u) \right)
                      + \frac{\epsilon (t-u)}{s^3}
                        \left( I_3^{d=6-2\epsilon}(t)
                             - I_3^{d=6-2\epsilon}(u) \right) & \\
                      + \frac{\epsilon(1-\epsilon)}{s^2}
                        \left( I_4^{d=8-2\epsilon}(s,t)
                             + I_4^{d=8-2\epsilon}(s,u)
                             + I_4^{d=8-2\epsilon}(t,u) \right) &
                  \bigg\} \,,
      \label{N4gravitymatterall}
	\end{aligned} \end{equation}
which is valid to all orders in $\epsilon$.

      All of one-loop constructions with $\cN>4$,
as we discuss in section~\ref{sec:comparison},
fit automatically in the string-theoretic picture of the BCJ double copy.
This is due to fact that, just as the field-theoretic numerators are
independent of the loop momentum,
the $\cN=4$ string-based integrands do not depend on Schwinger proper times.

\subsection{Double copy of one $\cN=2$ SYM}
\label{sec:ft-N2}

      The second option to compute
$\mathcal{M}_{\cN=4,\text{matt}}^{\text{1-loop}}(1^-,2^-,3^+,4^+)$
requires the BCJ representation for the $\cN\!=\!2$ hyper-multiplet amplitude.
The latter can also be used to construct gravity amplitudes
with $\cN<4$ supersymmetries \cite{Carrasco:2012ca},
such as
$(\cN\!=\!1\text{ gravity}) = (\cN\!=\!1\text{ SYM})
                       \times (\text{pure Yang-Mills})$.
However, we will consider it mostly in the context of obtaining
the BCJ numerators for $\cN\!=\!2$ SYM:
\begin{equation}
      n_{\cN=2,\text{vect}} (\ell) =
      n_{\cN=4,\text{vect}} - 2 \,
      n_{\cN=2,\text{hyper}} (\ell) \,,
\label{numeratorN2}
\end{equation}
whose double-copy square $\cN\!=\!4$ supergravity coupled to two $\cN\!=\!4$
matter multiplets:
      \begin{equation}
            \mathcal{M}_{(\cN=2)\times(\cN=2),\text{grav}}^{\text{1-loop}} =
            \mathcal{M}_{\cN=8,\text{grav}}^{\text{1-loop}}
       - 4\,\mathcal{M}_{\cN=6,\text{matt}}^{\text{1-loop}}
       + 4\,\mathcal{M}_{\cN=4,\text{matt}}^{\text{1-loop}}  = 
            \mathcal{M}_{\cN=4,\text{grav}}^{\text{1-loop}}
       + 2\,\mathcal{M}_{\cN=4,\text{matt}}^{\text{1-loop}} \,.
      \label{N22expansion}
	\end{equation}
As a side comment, the problem of decoupling matter fields in this context
is analogous to the more difficult issue of constructing pure gravity
as a double copy of pure Yang-Mills~\cite{Johansson:2014zca}.

      Most importantly for the purposes of this paper,
$\cA_{\cN=2,\text{hyper}}^{\text{1-loop}}$ is the simplest four-point amplitude
with non-trivial loop-momentum dependence of the numerators, i.e. $O(\ell^2)$,
which is already reflected in its non-BCJ form~\eqref{N1chiral}
by the fact that no rational part is present in the integrated amplitudes.
The rest of this paper is mostly dedicated to studying both from the BCJ
construction and field theory the double copy
\begin{equation}
      (\cN\!=\!4\text{ matter}) = (\cN\!=\!2\text{ hyper})^2 \,.
\label{N22hyper}
\end{equation}
Here, the left-hand side stands for the contribution of vector matter multiplets
running in a four-graviton loop in $\cN\!=\!4$ supergravity,
while the right-hand side indicates multiplets running in a
four-gluon loop in SYM.
In the rest of this section, we obtain the field-theoretic numerators
for the latter amplitude contribution.
In the literature \cite{Bern:1994zx,Bern:1994cg,Carrasco:2012ca,Nohle:2013bfa},
it is also referred to as the contribution of the $\cN\!=\!1$ chiral multiplet
in the adjoint representation and is not to be confused with
the $\cN\!=\!1$ chiral multiplet in the fundamental representation,
the calculation for which can be found in \cite{Johansson:2014zca}.
By calling the former multiplet $\cN\!=\!2$ hyper, we avoid that ambiguity
and keep the effective number of supersymmetries explicit.

\subsection{Ansatz approach}
\label{sec:ansatz}

      The standard approach to finding kinematic numerators
which satisfy Jacobi identities is through an ansatz
\cite{Carrasco:2011mn,Carrasco:2012ca},
as to our knowledge, there is no general constructive way of doing this,
apart from the case of $\cN=4$ SYM at one loop \cite{Bjerrum-Bohr:2013iza}.
Recently, however, there has been considerable progress
\cite{Carrasco:2012ca,Chiodaroli:2013upa}
in applying orbifold constructions to finding BCJ numerators.

      In \cite{Carrasco:2012ca,Nohle:2013bfa,Chiodaroli:2013upa}, several types of
ansatz were used for one-loop four-point computations,
starting from three independent master box numerators $n^{\text{box}}(1^-2^-3^+4^+)$,
$n^{\text{box}}(1^-2^-4^+3^+)$ and $n^{\text{box}}(1^-4^+2^-3^+)$
from which all other cubic diagrams were constructed through Jacobi identities.

      In comparison, our ansatz starts with two master
boxes, $n^{\text{box}}(1^-,2^-,3^+,4^+)$ and $n^{\text{box}}(1^-,4^+,2^-,3^+)$,
considered as independent while $n^{\text{box}}(1^-,2^-,4^+,3^+)$
is obtained from $n^{\text{box}}(1^-,2^-,3^+,4^+)$ by exchanging momenta $k_3$
and $k_4$.

      From Feynman-rules power-counting, string theory and supersymmetry cancellations
\cite{Bern:1994cg} we expect numerators to have at most two powers of the loop momentum.
Moreover, the denominator of \eqref{N1chiral} contains $s$ and $s^2$,
but only $t$ and $u$.
Thus, it is natural to consider the following minimal ansatz:
      \begin{subequations} \begin{align}
            n_{\cN=2,\text{hyper}}^{\text{box}} &
                  (1^-,2^-,3^+,4^+;\ell)
                  = \, i st A^{tree}(1^-,2^-,3^+,4^+) \\ \times
                  \frac{1}{s^2 t u} &
                  \bigg\{
                        P^{\text{(split-hel)}}_{4;2;1}
                        \left( s,t; (\ell\!\cdot\!k_1),(\ell\!\cdot\!k_2),
                                    (\ell\!\cdot\!k_3);\ell^2 \right)
                      + P^{\text{(split-hel)}}_{2}(s,t) \,
                        4 i \epsilon(k_1,k_2,k_3,\ell)
                  \bigg\} \,, \nonumber \\
            n_{\cN=2,\text{hyper}}^{\text{box}} &
                  (1^-,4^+,2^-,3^+;\ell)
                  = \, i st A^{tree}(1^-,2^-,3^+,4^+) \\ \times
                  \frac{1}{s^2 t u} &
                  \bigg\{
                        P^{\text{(alt-hel)}}_{4;2;1}
                        \left( s,t; (\ell\!\cdot\!k_1),(\ell\!\cdot\!k_2),
                                    (\ell\!\cdot\!k_3);\ell^2 \right)
                      + P^{\text{(alt-hel)}}_{2}(s,t) \,
                        4 i \epsilon(k_1,k_2,k_3,\ell)
                  \bigg\} \,. \nonumber
	\end{align} \label{N1ansatz} \end{subequations}

In eq.~\eqref{N1ansatz}, $P_{2}(s,t)$ is a homogeneous polynomial of degree 2
and $P_{4;2;1} \left( s, t; \tau_1, \tau_2, \tau_3; \lambda \right)$
is a homogeneous polynomial of degree 4, but not greater than 2 for arguments
$\tau_1$, $\tau_2$ and $\tau_3$ and at most linear in the last argument $\lambda$.
The 84 coefficients of these polynomials are the free parameters of the ansatz,
that we shall determine from the kinematic Jacobi identities and cut constraints.

Following \cite{Carrasco:2012ca}, we introduced in \eqref{N1ansatz} parity-odd terms
      \begin{equation}
            \epsilon(k_1,k_2,k_3,\ell) = \epsilon_{\lambda \mu \nu \rho}
                     k_1^{\lambda} k_2^{\mu} k_3^{\nu} \ell^{\rho} \,,
      \label{epsilon}
	\end{equation}
which integrate to zero in gauge theory
but may contribute to gravity when squared in the double copy construction.

      \begin{figure}[t]
      \centering
      \begin{subfigure}[b]{0.75\textwidth}
      \parbox{90pt}{ \begin{fmffile}{graph1}
      \fmfframe(10,10)(10,10){ \begin{fmfgraph*}(60,48)
            \fmflabel{$1^-$}{g1}
            \fmflabel{$2^-$}{g2}
            \fmflabel{$3^+$}{g3}
            \fmflabel{$4^+$}{g4}
            \fmfleft{g1,g2}
            \fmfright{g4,g3}
            \fmf{plain}{g1,v1}
            \fmf{plain}{g2,v2}
            \fmf{plain}{g3,v3}
            \fmf{plain}{g4,v4}
            \fmf{plain,tension=0.20}{v1,v2}
            \fmf{plain,tension=0.20}{v2,v3}
            \fmf{plain,tension=0.20}{v3,v4}
            \fmf{plain_arrow,tension=0.20,label=$\ell$,label.side=left}{v4,v1}
      \end{fmfgraph*} }
      \end{fmffile} }
      $ = \quad $
      \parbox{90pt}{ \begin{fmffile}{graph2}
      \fmfframe(10,10)(10,10){ \begin{fmfgraph*}(60,48)
            \fmflabel{$1^-$}{g1}
            \fmflabel{$2^-$}{g2}
            \fmflabel{$3^+$}{g3}
            \fmflabel{$4^+$}{g4}
            \fmfleft{g2,g1}
            \fmfright{g3,g4}
            \fmf{plain}{g1,v1}
            \fmf{plain}{g2,v2}
            \fmf{plain}{g3,v3}
            \fmf{plain}{g4,v4}
            \fmf{plain,tension=0.20}{v1,v2}
            \fmf{plain_arrow,tension=0.20,label=$k_1\!+\!k_2\!-\!\ell$,
                          label.dist=10pt,label.side=left}{v3,v2}
            \fmf{plain,tension=0.20}{v3,v4}
            \fmf{plain,tension=0.20}{v4,v1}
      \end{fmfgraph*} }
      \end{fmffile} }
      \end{subfigure}

      \vspace{1em}

      \begin{subfigure}[b]{0.75\textwidth}
      \parbox{90pt}{ \begin{fmffile}{graph3}
      \fmfframe(10,10)(10,10){ \begin{fmfgraph*}(60,48)
            \fmflabel{$1^-$}{g1}
            \fmflabel{$4^+$}{g4}
            \fmflabel{$2^-$}{g2}
            \fmflabel{$3^+$}{g3}
            \fmfleft{g1,g4}
            \fmfright{g3,g2}
            \fmf{plain}{g1,v1}
            \fmf{plain}{g2,v2}
            \fmf{plain}{g3,v3}
            \fmf{plain}{g4,v4}
            \fmf{plain,tension=0.20}{v1,v4}
            \fmf{plain,tension=0.20}{v4,v2}
            \fmf{plain,tension=0.20}{v2,v3}
            \fmf{plain_arrow,tension=0.20,label=$\ell$,label.side=left}{v3,v1}
      \end{fmfgraph*} }
      \end{fmffile} }
      $ = \quad $
      \parbox{90pt}{ \begin{fmffile}{graph4}
      \fmfframe(10,10)(10,10){ \begin{fmfgraph*}(60,48)
            \fmflabel{$1^-$}{g1}
            \fmflabel{$4^+$}{g4}
            \fmflabel{$2^-$}{g2}
            \fmflabel{$3^+$}{g3}
            \fmfleft{g1,g3}
            \fmfright{g4,g2}
            \fmf{plain}{g1,v1}
            \fmf{plain}{g2,v2}
            \fmf{plain}{g3,v3}
            \fmf{plain}{g4,v4}
            \fmf{plain_arrow,tension=0.20,label=$k_1\!-\!\ell$,
                                          label.side=left}{v4,v1}
            \fmf{plain,tension=0.20}{v4,v2}
            \fmf{plain,tension=0.20}{v2,v3}
            \fmf{plain,tension=0.20}{v3,v1}
      \end{fmfgraph*} }
      \end{fmffile} }
      $ = \quad $
      \parbox{90pt}{ \begin{fmffile}{graph5}
      \fmfframe(10,10)(10,10){ \begin{fmfgraph*}(60,48)
            \fmflabel{$1^-$}{g1}
            \fmflabel{$4^+$}{g4}
            \fmflabel{$2^-$}{g2}
            \fmflabel{$3^+$}{g3}
            \fmfleft{g2,g4}
            \fmfright{g3,g1}
            \fmf{plain}{g1,v1}
            \fmf{plain}{g2,v2}
            \fmf{plain}{g3,v3}
            \fmf{plain}{g4,v4}
            \fmf{plain,tension=0.20}{v1,v4}
            \fmf{plain,tension=0.20}{v4,v2}
            \fmf{plain_arrow,tension=0.20,label=$-\ell\!-\!k_3$,
                                          label.side=left}{v3,v2}
            \fmf{plain,tension=0.20}{v3,v1}
      \end{fmfgraph*} }
      \end{fmffile} }
      \end{subfigure}
      \caption{Box graph symmetries \label{boxsymmetries}}
      \end{figure}
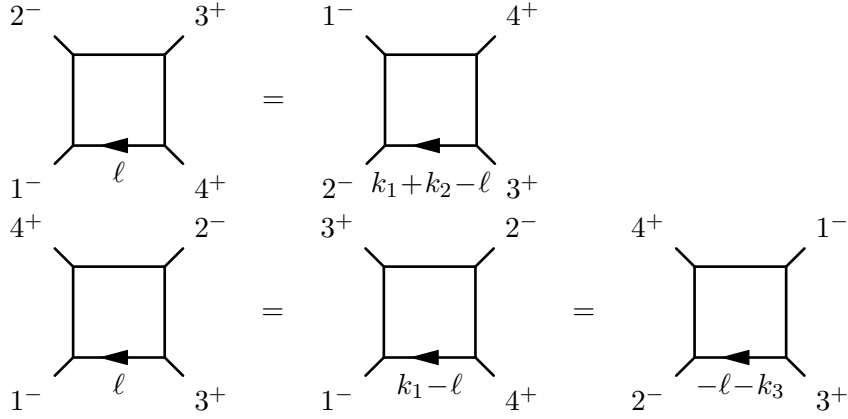
      The first constraints on the coefficients of the ansatz come from imposing
the obvious graph symmetries shown in figure~\ref{boxsymmetries} given by
      \begin{equation} \begin{aligned}
            n^{\text{box}}(1^-,2^-,3^+,4^+; \ell)
        & = n^{\text{box}}(2^-,1^-,4^+,3^+; k_1\!+\!k_2\!-\!\ell) \,, \\
            n^{\text{box}}(1^-,4^+,2^-,3^+; \ell)
        & = n^{\text{box}}(1^-,3^+,2^-,4^+; k_1\!-\!\ell)
          = n^{\text{box}}(2^-,4^+,1^-,3^+; -\ell\!-\!k_3) \,,
	\end{aligned} \label{boxflip} \end{equation}
after which 45 coefficients remain unfixed.

      Another set of constraints comes from the cuts.
In particular, the quadruple cuts provide 10 more constraints on the master boxes alone.
As we define triangle and bubble numerators through numerator Jacobi identities,
such as the one shown in figure~\ref{fig:jacobi1}, 35 remaining parameters
propagate to other numerators and then define the full one-loop integrand.
Note that whenever there are multiple Jacobi identities
defining one non-master numerator,
the master graph symmetries (\ref{boxflip}) guarantee that they are equivalent.

      Double cuts are sensitive not only to boxes, but also to triangles and
bubbles.
Imposing them gives further 18 constraints.
As a consistency check, we can impose double cuts without imposing
quadruple cuts beforehand
and in that case double cuts provide 28 conditions
with the 10 quadruple-cut constraints being a subset of those.
In any case, we are left with 17 free parameters after imposing all physical
cuts.

      For simplicity, we choose to impose another set of conditions:
vanishing of all bubble numerators,
including bubbles on external lines
(otherwise rather ill-defined in the massless case).
This is consistent with the absence of bubbles in our string-theoretic setup
of section~\ref{sec:st-loop}.
Due to sufficiently high level of supersymmetry,
that does not contradict the cuts and helps us eliminate
14 out of 17 free coefficients.
Let us call 3 remaining free coefficients $\As$, $\Ae$ and $\At$.
For any values of those,
we can check by direct computation that our solution integrates to
\eqref{N1chiral},
which is a consequence of the cut-constructibility
of supersymmetric gauge theory amplitudes.
However, there is still one missing condition,
which we will find from the $d$-dimensional cuts in section~\ref{sec:ft-gravity}.

\subsection{Double copy and $d$-dimensional cuts}
\label{sec:ft-gravity}

The double copy of the gluon amplitude with $\cN=2$ hyper multiplet in the loop
naturally produces the graviton amplitude with $\cN=4$ matter multiplet in the loop,
as in \eqref{N22hyper}.
First, we check that the gravity integrand satisfies all cuts.

So far we have been considering only four-dimensional cuts
and cut-constructible gauge theory amplitudes for which it does not matter
if during integration $\ell^2$ term in the numerator is considered as
4- or $(4-2\epsilon)$-dimensional.
After all, the difference will be just $\mu^2 = \ell^2_{(4)} - \ell^2_{(d)} $
which integrates to $O(\epsilon)$.
Note that we consider external momenta to be strictly four-dimensional, thus
the scalar products with external momenta $k_i$ like $\ell_{(d)}\cdot
k_i=\ell_{(4)}\cdot k_i$ are four-dimensional automatically.

The issue is that now $\cN=4$ gravity amplitudes are not longer
cut-constructible,
so the fact that double copy satisfies all four-dimensional cuts is not enough
to guarantee the right answer.
This is reflected by the fact that the difference
between $\ell^4_{(4)}$ and $\ell^4_{(d)}$ now integrates to $O(1)$
and produces rational terms.
It seems natural to treat $\ell$ in (\ref{N1ansatz}) as strictly
four-dimensional.
Then our gravity solution integrates to
      \begin{equation} \begin{aligned}           
            \mathcal{M}_{\cN=4,\text{matt}}^{\text{1-loop}}(1^-,2^-,3^+,4^+)
                  = \braket{12}^4 [34]^4 \frac{i r_{\Gamma}}{ (4\pi)^{d/2} }
                  \frac{1}{2 s^4}
                  \bigg\{
                    \!& - t u \left( \ln^2\!\left( \frac{-t}{-u} \right)
                                   + \pi^2 \right) \\
                        + s(t-u) \ln\!\left( \frac{-t}{-u} \right)
                      & + s^2 \left( 1 - \frac{1}{16} (3 + 2 \At)^2 \right)
                  \bigg\} \,,
      \label{N4gravitymatter}
	\end{aligned} \end{equation}
where $r_{\Gamma}$ the standard prefactor defined in \eqref{rgamma}.
That coincides with the known answer from \cite{Dunbar:1994bn}
and the truncated version of \eqref{N4gravitymatterall} \cite{Bern:2011rj},
if $ \At = -3/2 $.

      \begin{figure}[t]
      \centering
      \parbox{127pt}{ \begin{fmffile}{graph20}
      \fmfframe(10,10)(10,10){ \begin{fmfgraph*}(100,50)
            \fmflabel{$1^-$}{g1}
            \fmflabel{$2^-$}{g2}
            \fmflabel{$3^+$}{g3}
            \fmflabel{$4^+$}{g4}
            \fmfleft{g1,g2}
            \fmfright{g4,g3}
            \fmftop{top}
            \fmfbottom{bottom}
            \fmf{gluon}{g1,vleft}
            \fmf{gluon}{g2,vleft}
            \fmf{gluon}{g3,vright}
            \fmf{gluon}{g4,vright}
            \fmf{dashes}{top,bottom}
            \fmf{plain,left,tension=0.55}{vleft,vright}
            \fmf{plain,right,tension=0.55}{vleft,vright}
            \fmfblob{0.17w}{vleft}
            \fmfblob{0.17w}{vright}
      \end{fmfgraph*} }
      \end{fmffile} }
      \caption{$s$-channel cut for
               $ A_{\cN=2,\text{hyper}}^{\text{1-loop}}(1^-,2^-,3^+,4^+) $
               \label{Scut}}
      \end{figure}
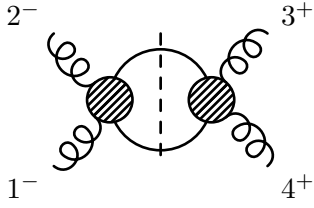

For the double copy to have predictive power beyond the cut-constructible cases,
one should start with gauge theory numerators that satisfy all $d$-dimensional cuts.
For $\cN=2$ SYM, the difference should just be related to
the $\mu^2$ ambiguity mentioned above.
As we already know that we miss only one extra condition,
it suffices to consider the simplest cut sensitive to $\mu^2$ terms,
i.e. the $s$-channel cut
for $ A_{\cN=2,\text{hyper}}^{\text{1-loop}}(1^-,2^-,3^+,4^+) $
that vanishes in four dimensions (figure~\ref{Scut}).

We can either construct this cut from massive scalar and fermion amplitudes
provided in \cite{Bern:1995db},
or simply use their final $d$-dimensional expression for this color-ordered amplitude:
      \begin{equation} \begin{aligned}
            A_{\cN=2,\text{hyper}}^{\text{1-loop}}(1^-,2^-,3^+,4^+) =
                  \braket{12}^2 [34]^2 \frac{i}{ (4\pi)^{d/2} }
                  \bigg\{&
                      - \frac{1}{st} I_2(t)
                      + \frac{1}{s} I_4(s,t)[\mu^2]
                  \bigg\} \,.
      \label{N1chiralFull}
      \end{aligned} \end{equation}
Unifying all our gauge theory numerators into one box and making use of
massive $s$-cut kinematics we retrieve the the following cut expression:
      \begin{equation} \begin{aligned}
            \delta_s
            A_{\cN=2,\text{hyper}}^{\text{1-loop}}(1^-,2^-,3^+,4^+) =
                  \braket{12}^2 [34]^2
                  \frac{1}{s}
                  \int \! d\text{LIPS}
                  \frac{ \mu^2 \left(1 - \frac{1}{4}(3+2\At) \right) }
                       { ((\ell-k_1)-\mu^2) ((\ell+k_4)-\mu^2) } \,,
      \label{N1chiralScut}
      \end{aligned} \end{equation}
which coincides with the $s$-cut of (\ref{N1chiralFull}) if $ \At = -3/2 $.
Thus, we have reproduced the missing condition invisible to four-dimensional cuts.

We preserve the remaining two-parameter freedom and write down
the full set of numerators for the $\cN=2$ hyper
(or, equivalently, $\cN=1$ chiral) multiplet amplitude as follows:
      \begin{subequations} \begin{align}
      \parbox{70pt}{ \begin{fmffile}{graph12}
      \fmfframe(0,7)(0,46){ \begin{fmfgraph*}(60,45)
            \fmflabel{$1^-$}{g1}
            \fmflabel{$2^-$}{g2}
            \fmflabel{$3^+$}{g3}
            \fmflabel{$4^+$}{g4}
            \fmfleft{g1,g2}
            \fmfright{g4,g3}
            \fmf{plain}{g1,v1}
            \fmf{plain}{g2,v2}
            \fmf{plain}{g3,v3}
            \fmf{plain}{g4,v4}
            \fmf{plain,tension=0.20}{v1,v2}
            \fmf{plain,tension=0.20}{v2,v3}
            \fmf{plain,tension=0.20}{v3,v4}
            \fmf{plain_arrow,tension=0.20,label=$\ell$,
                                          label.side=left}{v4,v1}
      \end{fmfgraph*} }
      \end{fmffile} }
      &
      \begin{aligned}
        =\,& \frac{\ell^2}{s} - \frac{(\ell(k_1+k_3))}{2 u}
          + \frac{1}{s^2} [ s (\ell k_1) - t (\ell k_2) + u (\ell k_3) ] \\
        & - \frac{1}{s^2} \!
            \left[ (\ell k_1)^2 \!+ (\ell k_2)^2 \!+ 6 (\ell k_1) (\ell k_2)
                 - t \frac{(\ell(k_1+k_3))^2}{u} - u \frac{(\ell(k_2+k_3))^2}{t}
            \right] \\
        & + \frac{\Ae}{s^2} \,
            [ (\ell k_1)^2 + (\ell k_2)^2
            - (\ell k_3)^2 - (\ell k_4)^2 + u (\ell(k_2+k_3)) ] \\
        & + \frac{\As (s^2 + t u)}{s^2 t u}
            (\ell(k_2+k_3)) [ 2 (\ell(k_1+k_3)) - u ] \,,
           \label{graph:box1234gauge}
      \end{aligned}
      \\
      \parbox{70pt}{ \begin{fmffile}{graph13}
      \fmfframe(0,7)(0,44){ \begin{fmfgraph*}(60,45)
            \fmflabel{$1^-$}{g1}
            \fmflabel{$4^+$}{g4}
            \fmflabel{$2^-$}{g2}
            \fmflabel{$3^+$}{g3}
            \fmfleft{g1,g4}
            \fmfright{g3,g2}
            \fmf{plain}{g1,v1}
            \fmf{plain}{g2,v2}
            \fmf{plain}{g3,v3}
            \fmf{plain}{g4,v4}
            \fmf{plain,tension=0.20}{v1,v4}
            \fmf{plain,tension=0.20}{v4,v2}
            \fmf{plain,tension=0.20}{v2,v3}
            \fmf{plain_arrow,tension=0.20,label=$\ell$,
                                          label.side=left}{v3,v1}
      \end{fmfgraph*} }
      \end{fmffile} }
      &
      \begin{aligned}
        =\,& \frac{\ell^2}{s} + \frac{(\ell(k_1+k_2))}{2 s}
          - \frac{t (\ell(k_1+k_3))}{s^2} - \frac{2i \epsilon(k_1,k_2,k_3,\ell)}{s^2} \\
        & - \frac{1}{s^2} \!
            \left[ (\ell k_1)^2 \!+ (\ell k_2)^2 \!+ 6 (\ell k_1) (\ell k_2)
                 - t \frac{(\ell(k_1+k_3))^2}{u} - u \frac{(\ell(k_2+k_3))^2}{t}
            \right] \\
        & + \frac{\Ae}{s^2} \,
            [ (\ell k_1)^2 + (\ell k_2)^2
            - (\ell k_3)^2 - (\ell k_4)^2 - t (\ell(k_1+k_3)) ] \\
        & + \frac{\As (s^2 + t u)}{s^2 t u}
            (\ell(k_1+k_3)) [ 2 (\ell(k_2+k_3)) + t ] \,,
           \label{graph:box1423gauge}
      \end{aligned}
	\end{align} \label{boxesfinal} \end{subequations}

      \begin{subequations} \begin{align}
      \parbox{70pt}{ \begin{fmffile}{graph14}
      \fmfframe(0,15)(0,15){ \begin{fmfgraph*}(60,45)
            \fmflabel{$1^-$}{g1}
            \fmflabel{$2^-$}{g2}
            \fmflabel{$3^+$}{g3}
            \fmflabel{$4^+$}{g4}
            \fmfleft{g1,g2}
            \fmfright{g4,g3}
            \fmf{plain,tension=0.60}{g1,v12,g2}
            \fmf{plain}{g3,v3}
            \fmf{plain}{g4,v4}
            \fmf{plain,tension=0.60}{v12,v34}
            \fmf{plain,tension=0.30}{v34,v3}
            \fmf{plain,tension=0.30}{v3,v4}
            \fmf{plain_arrow,tension=0.30,label=$\ell$,
                                          label.side=left}{v4,v34}
      \end{fmfgraph*} }
      \end{fmffile} }
      &
      \begin{aligned}
         = 
           \left(
              - \frac{1 - \As + \Ae}{s^2}
              - \frac{1 - 2 \As}{2 t u}
           \right)
           [ s (\ell k_2) - t (\ell k_4) - u (\ell k_3) ] \,,
      \end{aligned}
      \\
      \parbox{70pt}{ \begin{fmffile}{graph15}
      \fmfframe(0,15)(0,15){ \begin{fmfgraph*}(60,45)
            \fmflabel{$1^-$}{g1}
            \fmflabel{$2^-$}{g2}
            \fmflabel{$3^+$}{g3}
            \fmflabel{$4^+$}{g4}
            \fmfleft{g1,g2}
            \fmfright{g4,g3}
            \fmf{plain}{g1,v1}
            \fmf{plain}{g2,v2}
            \fmf{plain,tension=0.60}{g3,v34,g4}
            \fmf{plain,tension=0.60}{v34,v12}
            \fmf{plain,tension=0.30}{v1,v2}
            \fmf{plain,tension=0.30}{v2,v12}
            \fmf{plain_arrow,tension=0.30,label=$\ell$,
                                          label.side=left}{v12,v1}
      \end{fmfgraph*} }
      \end{fmffile} }
      &
      \begin{aligned}
         = 
           \left(
              - \frac{1 - \As + \Ae}{s^2}
              + \frac{1 + 2 \As}{2 t u}
           \right)
           [ s (\ell k_3) - t (\ell k_1) - u (\ell k_2) ] \,,
      \end{aligned}
      \\
      \parbox{70pt}{ \begin{fmffile}{graph16}
      \fmfframe(0,15)(0,15){ \begin{fmfgraph*}(60,60)
            \fmflabel{$1^-$}{g1}
            \fmflabel{$2^-$}{g2}
            \fmflabel{$3^+$}{g3}
            \fmflabel{$4^+$}{g4}
            \fmfleft{g1,g2}
            \fmfright{g4,g3}
            \fmf{plain,tension=0.60}{g1,v14,g4}
            \fmf{plain}{g2,v2}
            \fmf{plain}{g3,v3}
            \fmf{plain,tension=0.60}{v14,v23}
            \fmf{plain,tension=0.30}{v2,v3}
            \fmf{plain,tension=0.30}{v3,v23}
            \fmf{plain_arrow,tension=0.30,label=$\ell\!-\!k_1$,
                                          label.side=left}{v23,v2}
      \end{fmfgraph*} }
      \end{fmffile} }
      &
      \begin{aligned} \label{triangleexample}
         = 
                \frac{1}{2 s u}
                [su - s (\ell k_3) + t (\ell k_1) - u (\ell k_2) ]
              + \frac{2i \epsilon(k_1,k_2,k_3,\ell)}{s^2} \,,
      \end{aligned}
      \\
      \parbox{70pt}{ \begin{fmffile}{graph17}
      \fmfframe(0,15)(0,15){ \begin{fmfgraph*}(60,60)
            \fmflabel{$1^-$}{g1}
            \fmflabel{$2^-$}{g2}
            \fmflabel{$3^+$}{g3}
            \fmflabel{$4^+$}{g4}
            \fmfleft{g1,g2}
            \fmfright{g4,g3}
            \fmf{plain,tension=0.60}{g2,v23,g3}
            \fmf{plain}{g4,v4}
            \fmf{plain}{g1,v1}
            \fmf{plain,tension=0.60}{v23,v14}
            \fmf{plain,tension=0.30}{v1,v14}
            \fmf{plain,tension=0.30}{v14,v4}
            \fmf{plain_arrow,tension=0.30,label=$\ell$,
                                          label.side=left}{v4,v1}
      \end{fmfgraph*} }
      \end{fmffile} }
      &
      \begin{aligned}
         = 
              - \frac{1}{2 s u}
                [ s (\ell k_1) - t (\ell k_3) + u (\ell k_4) ]
              - \frac{2i \epsilon(k_1,k_2,k_3,\ell)}{s^2} \,,
      \end{aligned}
      \\
      \parbox{70pt}{ \begin{fmffile}{graph18}
      \fmfframe(0,15)(0,15){ \begin{fmfgraph*}(60,60)
            \fmflabel{$1^-$}{g1}
            \fmflabel{$4^+$}{g4}
            \fmflabel{$2^-$}{g2}
            \fmflabel{$3^+$}{g3}
            \fmfleft{g1,g4}
            \fmfright{g3,g2}
            \fmf{plain,tension=0.60}{g1,v13,g3}
            \fmf{plain}{g2,v2}
            \fmf{plain}{g4,v4}
            \fmf{plain,tension=0.60}{v13,v24}
            \fmf{plain,tension=0.30}{v4,v2}
            \fmf{plain,tension=0.30}{v2,v24}
            \fmf{plain_arrow,tension=0.30,label=$\ell\!-\!k_1$,
                                          label.side=left}{v24,v4}
      \end{fmfgraph*} }
      \end{fmffile} }
      &
      \begin{aligned}
         = 
                \frac{1}{2 s t}
                [ st + s (\ell k_2) + t (\ell k_4) - u (\ell k_3) ]
              - \frac{2i \epsilon(k_1,k_2,k_3,\ell)}{s^2} \,,
      \end{aligned}
      \\
      \parbox{70pt}{ \begin{fmffile}{graph19}
      \fmfframe(0,15)(0,15){ \begin{fmfgraph*}(60,60)
            \fmflabel{$1^-$}{g1}
            \fmflabel{$4^+$}{g4}
            \fmflabel{$2^-$}{g2}
            \fmflabel{$3^+$}{g3}
            \fmfleft{g1,g4}
            \fmfright{g3,g2}
            \fmf{plain,tension=0.60}{g2,v24,g4}
            \fmf{plain}{g1,v1}
            \fmf{plain}{g3,v3}
            \fmf{plain,tension=0.60}{v24,v13}
            \fmf{plain,tension=0.30}{v1,v13}
            \fmf{plain,tension=0.30}{v13,v3}
            \fmf{plain_arrow,tension=0.30,label=$\ell$,
                                          label.side=left}{v3,v1}
      \end{fmfgraph*} }
      \end{fmffile} }
      &
      \begin{aligned}
         = 
              - \frac{1}{2 s t}
                [ s (\ell k_3) + t (\ell k_1) - u (\ell k_2) ]
              - \frac{2i \epsilon(k_1,k_2,k_3,\ell)}{s^2} \,,
      \end{aligned}
      \end{align} \label{trianglesfinal} \end{subequations}
where for brevity we omitted the trivial kinematic prefactor $\braket{12}^2[34]^2$.

The numerators that we obtain are non-local,
as they contain inverse powers of Mandelstam invariants
on top of those already included in their denominators.
This is a feature of using the spinor-helicity formalism for BCJ numerators
\cite{Broedel:2011pd,Carrasco:2011mn,Carrasco:2012ca,Chiodaroli:2013upa}
and is understood to be due to the choice of helicity states for the external gluons.
Indeed, the numerators given in \cite{Bern:2013yya}
in terms of polarization vectors are local while gauge-dependent.

We first note that the box numerators~\eqref{boxesfinal}
do not possess constant terms. Later, we will relate this to a similar
absence of constant terms in the string-based integrand.
Moreover, the triangles integrate to null contributions
to gauge theory amplitudes~\eqref{N1chiral}.
Nonetheless, they are necessary for the double copy construction
of the gravity amplitude~\eqref{N4gravitymatter},
where they turn out to integrate to purely six-dimensional scalar triangles
$I_3^{d=6-2\epsilon}$.
The easiest way to check these statements is to explicitly convert
the triangle numerators~\eqref{trianglesfinal} to the Feynman parameter space,
as explained in appendix~\ref{app:triangles}.
We will use both of these facts later in section~\ref{sec:comparison}.

Finally, there are conjugation relations that hold for the final amplitudes,
but are not automatic for the integrand numerators:
      \begin{subequations} \begin{align}
         \left(
         n_{\cN=2,\text{hyper}}^{\text{box}}(1^-,2^-,3^+,4^+; \ell) \right)^* & =
         n_{\cN=2,\text{hyper}}^{\text{box}}(4^-,3^-,2^+,1^+; -\ell) \,, \\
         \left(
         n_{\cN=2,\text{hyper}}^{\text{box}}(1^-,4^+,2^-,3^+; \ell) \right)^* & =
         n_{\cN=2,\text{hyper}}^{\text{box}}(3^-,2^+,4^-,1^+; -\ell) \,.
	\end{align} \label{conjugation} \end{subequations}
Although they are not necessary for the integrated results to be correct,
one might choose to enforce them at the integrand level,
which would fix both remaining parameters to
\begin{equation}
      \As=0 \,, \quad \Ae= -1 \,,
\label{e:A7A11conj}
\end{equation}
and thus produce the unambiguous bubble-free BCJ solution.
However, leaving two parameters unfixed can have its advantages to discern
analytically pure coincidences from systematic patterns at the integrand level.

\section{One loop in string theory}
\label{sec:st-loop}
This section is mostly a review of a detailed calculation given in
\cite{Tourkine:2012vx} in order to explain the string-theoretic origin
of the worldline integrands of the $\cN=2$ SYM and symmetric
$\cN=4$ supergravity, in heterotic string and type II string in $d=4-2\epsilon$
dimensions, respectively.

The reader not familiar with the worldline formalism may simply observe that
the general formula~\eqref{e:W-col-kin} contains a contribution to the gravity amplitude
which mixes the left- and the-right-moving sectors and thus makes it
look structurally different
from the double copy construction.
Then the $\cN\!=\!2$ gauge theory and the $\cN\!=\!4$ gravity integrands
are given in eqs.~\eqref{e:Neq2gauge} and \eqref{e:Neq4grav},
respectively, in terms of the Schwinger proper-time variables.
They are integrated according to \eqref{e:WX-def}.
These are the only building blocks needed to go directly to section~\ref{sec:comparison},
where the link between the worldline formalism and the usual Feynman diagrams
is described starting from the loop-momentum space.

\subsection{Field theory amplitudes from string theory}
\label{sec:limitloop}

A detailed set of rules known as the Bern-Kosower rules was developed
in~\BKcite~to compute gauge theory amplitudes from the field theory limit of
fermionic models
in heterotic string theory. It was later extended to asymmetric constructions
of supergravity amplitudes in \cite{Bern:1993wt,Dunbar:1994bn}
(see also the review \cite{Schubert:2001he} and the approach of
\cite{DiVecchia:1996uq,DiVecchia:1996kf,Magnea:2013lna} using the Schottky
parametrization).
One-loop amplitudes in the open string are known at any
multiplicity in the pure spinor formalism \cite{Mafra:2012kh}.

Here we recall the general mechanism to extract the field theory limit of string
amplitudes at one loop in the slightly different context of orbifold models of
the heterotic and type II string. A general four-point closed-string theory 
amplitude writes
\begin{equation}
  \mathcal A^{\rm{string}}_{\text{1-loop}} = N \int_\cF {\d^2\tau \over
\tau_2^2}\int_\cT\prod_{i=1}^3
{\d^2z_i\over\tau_2}\langle V_1(z_1) V_2(z_2) V_3(z_3) V_4(z_4) \rangle\,.
\label{e:one-loop-st}
\end{equation}
The integration domains are
$\cF=\{ \tau=\tau_1+i\tau_2; |\tau_1| \leq \frac12,|\tau|^2\geq1, \tau_2>0 \}$
and 
$\cT=\{ |\Re z| \leq \frac12,\, 0 \leq \Im z \leq \tau_2 \}$.
The normalization constant $N$ is different for heterotic and type II strings.
We will omit it throughout this section except the final formula \eqref{e:WX-def},
where the normalization is restored.
The $z_i$ are the positions of the vertex operators in the
complex torus $\cT$, and $z_4$ has been set to $z_4 = i\tau_2$ to fix
the genus-one conformal invariance.

On the torus, the fermionic fields $\psi^\mu$ and $\bar{\psi}^\nu$ can have
different boundary conditions when transported along the $A$- and $B$-cycles
(which correspond to the shifts $z\to z+1$ and $z\to z+\tau$, respectively). 
These boundary conditions define spin structures, denoted by two integers
$a,b \in \{0,1\}$ such that
\begin{equation}
   \psi(z+1)=e^{i\pi a}\psi(z)\,,\qquad
   \psi(z+\tau)=e^{i\pi b}\psi(z)\,.
\end{equation}
In an orbifold compactification, these boundary conditions can be mixed with
target-space shifts and the fields $X$ and $\psi$ can acquire
non-trivial boundary conditions, mixing the standard spin structures
(or Gliozzi-Scherk-Olive sectors) with more general orbifold sectors
\cite{Dixon:1985jw,Dixon:1986jc}.

The vertex operator correlation function \eqref{e:one-loop-st}
is computed in each orbifold and GSO sector,
using Wick's theorem with the two-point functions
\begin{subequations} \begin{align}
   \langle X(z) X(w) \rangle & = \cG(z-w,\tau) \label{e:torus-G} \,, \\
   \langle \psi(z) \psi(w) \rangle_{a,b} & = S_{a,b} (z-w,\tau) \label{e:torus-S} \,,
\end{align}  \label{e:torus-prop} \end{subequations}
whose explicit expressions are not needed for the purposes of this review.
They can be found, for instance, in \cite[eqs. (A.10),(A.19)]{Tourkine:2012vx}.
The vertex operators of the external gluons are in the $(0)$ picture
(see vertex operator obtained from \eqref{e:VOcurrent}$\times$\eqref{e:VOsusy0})
and the external gravitons ones are in the $(0,0)$ picture
(see \eqref{e:VOsusy0}$\times$\eqref{e:VOsusy0}). 

The total correlation function can then be written
in the following \emph{schematic} form:
\begin{equation}
 \mathcal A^{\rm string}_{\text{1-loop}} =
  \int_{\cF} \frac{\d^2\tau}{\tau_2^{d/2-3}} \,
  \int_{\cT} \prod_{i=1}^3 \frac{\d^2z_i}{\tau_2} \,\,
   \sum_{s,s'} \cZ^{s s^\prime} \!
   \left(\cW^{\rm (L)}_s(z) \ \overline{\cW_{s^\prime}^{\rm (R)}}(\bar z) 
       + \cW_{s,s'}^{\text{(L-R)}}(z,\bar z) \right) \, e^{\mathcal Q} \,.
\label{e:st-amp}
\end{equation}
where $s$ and $s^\prime$ correspond to the various GSO and orbifold sectors of
the theory with their corresponding conformal blocks,
and $\cZ^{s s^\prime}$ is defined so that it contains
the lattice factor $\Gamma_{10-d,10-d}$ or twistings thereof
according to the orbifold sectors and background Wilson lines.\footnote{More details
on these objects can be found in classical textbooks,
for instance, \cite[Chapters 9-10]{kiritsis2011string}.}
The exponent of the plane-wave factor $e^{\mathcal Q}$ writes explicitly
\begin{equation}
   \mathcal Q = \sum_{i<j} k_i \cdot k_j \,\, \cG(z_i-z_j) \,,
\label{e:Qdef}
\end{equation}
similarly to \eqref{e:KNtree} at tree level.
The partition function now intertwines the left- and right-moving CFT
conformal blocks and gives rise not only to pure chiral contractions
$\cW^{\rm (L)}_s \overline{\cW_{s'}^{\rm (R)}}$ 
but also to left-right mixed contractions $\cW_{s,s'}^{\text{(L-R)}}(z,\bar z)$.
The latter comes from terms such as
\begin{equation}
      \langle \partial X(z,\bar z) \bar \partial X(w,\bar w) \rangle 
      = -\ap \pi \delta^{(2)}(z-w) + \frac{\ap}{2\pi \Im \tau} \,,
\label{e:lr-one-loop}
\end{equation}
where the first term on the right-hand side gives a vanishing contribution
due the canceled propagator argument in the same way as at tree level.
The second term in eq.~\eqref{e:lr-one-loop} was absent at tree level,
see eq.~\eqref{e:lr-tree-level}, but now generates left-right mixed
contractions
in case the two sectors have coinciding target spaces, i.e. in gravity
amplitudes.
However, in gauge theory amplitudes in the heterotic string, the target
spaces are different, and contractions like~\eqref{e:lr-one-loop} do not occur.

The main computation that we use in this section was performed in great detail
in \cite{Tourkine:2012ip}, and the explicit expressions for partition functions,
lattice factors and conformal blocks may be found in the introductory sections thereof.

\begin{figure}[t]
      \centering
      \includegraphics[scale=1]{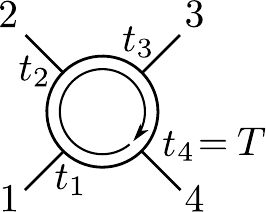}
\caption{Generic four-point worldline graph in the ordering $(1234)$.}
\label{fig:WL}
\end{figure}

The mechanism by which the string integrand descends to the worldline
(or tropical) integrand is qualitatively the same as at tree level.\footnote{
However, it has to be adapted to the geometry of the genus-one worldsheet.
In particular, phases of complex numbers on the sphere become
real parts of coordinates on the complex torus $\IC/(\IZ+\tau \IZ)$. 
As explained in \cite{Tourkine:2013rda}, this is due to the fact that the
complex torus is the image by the Abel-Jacobi map of the actual Riemann surface.
In the limit $\apt$, this map simplifies to a logarithmic map,
and coordinates on the surface $z,\bar z$ are related to coordinates on the
complex torus $\zeta,\bar\zeta$ by: $ \zeta=-2 i \pi \ap \ln z$. The same is
true for the modular parameter $\tau$, whose real and imaginary parts are
 linked to phases and modulus of $q$, respectively.}
In particular, one considers families of tori becoming thinner and thinner as
$\alpha'\to0$.
On these very long tori, only massless states can propagate
(massive states are projected out), so the level-matching condition of string theory
associated to the cylindrical coordinate on the torus can be integrated out
and the tori become worldloops.
Quantitatively, one performs the following well-known change of variables:
\begin{equation} \begin{aligned}
   \Im\tau \in \cF \quad& \longrightarrow \quad T = \Im\tau/\alpha' \in [0;+\infty[ \,,\\
   \Im z_i \in \cT \quad& \longrightarrow \quad t_i = z_i/\alpha' \in [0;T[ \,,
\end{aligned} \label{e:tropch-oneloop} \end{equation}
where $T$ is the Schwinger proper time of the loop\footnote{Strictly speaking,
as originally observed in \cite{Green:1999pv}
and recently reviewed in \cite{Tourkine:2013rda},
one should cut off the region $\cF$ by a parameter $L$,
so that the region of interest for us is actually the upper part $\Im\tau>L$ of $\cF$,
which in the field theory limit gives a hard Schwinger proper-time cutoff $T>\ap L$.
Here we trade this cutoff for dimensional regularization with $d=4-2\epsilon$.}
and $t_i$ are the proper times of the external legs along it (see figure~\ref{fig:WL}). 

We should also mention that to obtain a truly $d$-dimensional amplitude,
one should not forget to decouple Kaluza-Klein modes of the compactified string
by sending the radii $R$ of compactification to zero,
so that $R \sim \sqrt{\ap}$ (for instance,
in this way one sets the untwisted lattice factor $\Gamma_{10-d,10-d}$ to $1$).
The field theory worldline amplitude is obtained after that --- possibly lengthy
--- process of integrating out the real parts of $\tau$ and $z$'s, and one is
left with an integral of the form\footnote{Note that
the calligraphic letters $\cG,\cQ,\cW,\cZ$ refer to string-theoretic quantities,
whereas the plain letters $G,Q,W,Z$ refer to their worldline analogues.}
\cite{D'Hoker:1993ge}:
\begin{equation}
      \cA^{\text{1-loop}} =
      \int_0^\infty {\d T\over T^{d/2-3}}\,
      \int_0^1 ({\d u_i})^3\,\sum_{s,s'} Z_{s s'}
            \left( W^{\rm (L)}_s W^{\rm(R)}_{s'} + W^{\text{(L-R)}}_{s,s'} \right) \,
            e^{-T Q}\,,
\label{e:st-amp2}
\end{equation}
where $u_i=t_i/T$ are rescaled proper times.
As reviewed later in section~\ref{sec:comparison}, the exponential factor
$e^{-TQ}$
can also be regarded as a result of exponentiating the loop-momentum denominator
of the corresponding Feynman diagram,
with $Z_{s s'} \left( W^{\rm (L)}_s W^{\rm(R)}_{s'} + W^{\text{(L-R)}}_{s,s'} \right)$
coming from its numerator. 
Formula \eqref{e:st-amp2} can be written in terms of
derivatives of the worldline Green's function \cite{Strassler:1992zr,Dai:2006vj}
which descends from the worldsheet one and is defined by
\begin{equation}
   G(u_i,u_j)=T(|u_i-u_j|-(u_i-u_j)^2) \,.
\label{e:WLGF}
\end{equation}
For example, eq. \eqref{e:Qdef} becomes
\begin{equation}
   Q=\sum_{i<j} k_i \cdot k_j \,\, G(u_i,u_j) \,.
\label{e:Qtrop}
\end{equation}
The partition function factor $Z_{s s'}$, in the field theory limit,
just induces a sum over multiplet decompositions,
as in eqs.~\eqref{amplitudeN2}, \eqref{N4expansion} and \eqref{N6expansion},
but does not change the qualitative nature of the objects.

Moreover, it is worth mentioning that the field theory limit of mixed
contractions \eqref{e:lr-one-loop} produces only factors of $1/T$:
\begin{equation}
	\langle \partial X(z,\bar w) \bar \partial X(z,\bar w) \rangle
\underset{\apt}{\longrightarrow} -\frac{2}{T}
\label{e:lr-one-loop-worldline}
\end{equation}
without further dependence on the positions $t_i$ of the legs on the worldloop.
Note that in general, factors of $1/T^k$ modify the overall factor
$1/T^{d/2-(n-1)}$ and thus act as dimension shifts $d \to d+2k$.

Let us now discuss the differences between color and kinematics
in the integrand of eq.~\eqref{e:st-amp2}.
In heterotic string theory, the two sectors have different target spaces and
do not communicate with each other. In particular, the right-moving sector is
a color CFT: it is responsible for the color ordering in the field theory limit
as demonstrated in the Bern-Kosower papers~\BKcite, and its factor writes
\begin{equation}
   W^{\rm (R,\,color)} = \sum_{\mathfrak S\in S_{n-1}}
      \tr (T^{a_{\mathfrak S(1)}}...T^{a_{\mathfrak S(n-1)}}T^{a_n})
      \Theta(u_{\mathfrak S(1)}<...<u_{\mathfrak S(n-1)}<u_n) \,,
\end{equation}
where the sum runs over the set $S_{n-1}$ of permutations of $(n-1)$ elements.
It is multiplied by a $W^{\rm (L,\,\rm{kin})}$ which contains
the kinematical loop-momentum information. 

In gravity, both sectors are identical, they carry kinematical information and
can mix with each other. To sum up, we can write the following worldline formulas
for gauge theory and gravity amplitudes:
\begin{subequations} \begin{align}
  \cA_{\rm gauge}^{\text{1-loop}}\hspace{2pt} & = \int_0^\infty
  {\d T \over T^{d/2-3}}
  \, \int_0^1 \d^3u 
  \cdot \left( W^{\rm (L,\,kin)}\, W^{\rm (R,\,col)}\right)
  \cdot \,e^{-T Q}\,,
\label{e:Wgauge} \\
  \cM_{\rm gravity}^{\text{1-loop}} & = \int_0^\infty
  {\d T \over T^{d/2-3}}
  \, \int_0^1 \d^3u
  \cdot \left(W^{\rm (L,\,kin)}\,
  W^{\rm (R,\,kin)}+W^{\text{(L-R,\,kin)}}\right)
  \cdot \,e^{-T Q}\,.
\label{e:Wgravity}
\end{align} \label{e:W-col-kin} \end{subequations}
%
Besides the fact that these formulas are not written in the loop-momentum space,
the structure of the integrand of the gravity amplitude \eqref{e:Wgravity}
is different from the double-copy one in eq.~\eqref{e:Mgeneral}:
it has non-squared terms that come from left-right contractions. 
This paper is devoted to analysis of their role from the double copy point of view,
in the case of the four-point one-loop amplitude in
$(\cN\!=\!2)\times(\cN\!=\!2)$ gravity.

The kinematic correlators $W^{\rm kin}$ are always expressed as polynomials
in the derivatives of the worldline Green's function $G$:
\begin{equation}  \begin{aligned}
   \dot G(u_i,u_j)&={\rm sign}(u_i-u_j) - 2 (u_i-u_j) \,, \\
   \ddot G(u_i,u_j)&=\frac 2T \left(\delta(u_i-u_j) -1\right) \,,
\end{aligned} \label{e:Gwl} \end{equation}
where the factors of $T$ take into account the fact that the derivative is
actually taken with respect to the unscaled variables $t_i$,
where $\partial_{t_i} = T^{-1} \partial_{u_i} $.

To illustrate the link with the loop-momentum structure, let us recall the
qualitative dictionary between the worldline power-counting
and the loop-momentum one
\cite{BjerrumBohr:2008ji,BjerrumBohr:2008vc,Tourkine:2012ip}.
A monomial of the form $(\dot G)^n (\ddot G)^m$ contributes to $O(\ell^{n+2m})$
terms in four-dimensional loop-momentum integrals:\footnote{Therefore,
qualitatively, double derivatives count as squares of simple derivatives.
At one loop, an easy way to see this is to integrate by parts:
when the second derivative $\partial_{u_i}$ of $\ddot G(u_{ij})$
hits the exponential $e^{-T Q}$, a linear combination of $\dot G$ comes down
(see definition of $Q$ in eq. \eqref{e:Qtrop}) and produces $\dot G^2$.
In the non-trivial cases where one does not just have a single $\ddot G$ as a monomial,
it was proven \BKcite~that it is always possible to integrate out
all double derivatives after a finite number of integrations by parts.
Another possibility is to observe that the factor $1/T$ present in $\ddot G$
produces a dimension shift $d \to d+2$ in the worldline integrands,
which in terms of loop momentum schematically corresponds to
adding $\ell^2$ to the numerator of the $d$-dimensional integrand.}
\begin{equation}
   (\dot G)^n (\ddot G)^m \sim \ell^{n+2m}\,.
\label{e:dictionary}
\end{equation}
Later in section~\ref{sec:comparison}, we describe in more detail the converse relation,
i.e. the quantitative link between the loop momentum
and the Schwinger proper-time variables.

For definiteness, in order to have well-defined conventions for worldline integration,
we define a theory-dependent worldline numerator $W_X$
to be carrying  only loop-momentum-like information:
\begin{subequations} \begin{align}
   \cA_{X}^{\text{1-loop}}\hspace{2pt} \!=
      i\,\frac{2 t_8 F^4}{(4\pi)^{d/2}} \sum_{\mathfrak{S} \in S_3} &
      \tr(T^{a_{\mathfrak S(1)}} T^{a_{\mathfrak S(2)}} T^{a_{\mathfrak S(3)}} T^{a_4})
\label{e:WX-gauge} \\ \times &
      \int_0^\infty \!\!\! \frac{\d T}{T^{d/2-3}}
      \int_0^1 \!\!\! \d u_{\mathfrak{S}(1)}
      \int_0^{u_{\mathfrak{S}(1)}} \!\!\!\!\!\!\!\!\!\! \d u_{\mathfrak{S}(2)}
      \int_0^{u_{\mathfrak{S}(2)}} \!\!\!\!\!\!\!\!\!\! \d u_{\mathfrak{S}(3)}
         \cdot W_{X} \cdot e^{-T Q} \,, \nn \\
   \cM_{X}^{\text{1-loop}} = ~
      i\,\frac{4 t_8 t_8 R^4}{(4\pi)^{d/2}} ~~ &
      \int_0^\infty \!\!\! \frac{\d T}{T^{d/2-3}} \, \int_0^1 \!\!\!\d^3 u
         \cdot W_{X} \cdot e^{-T Q} \,.
\label{e:WX-grav}
\end{align} \label{e:WX-def} \end{subequations}
In \eqref{e:WX-gauge}, the sum runs over six orderings $\mathfrak{S}\in S_3$,
three out of which, $(123),(231),(312)$, are inequivalent
and correspond to the three kinematic channels $(s,t),\, (t,u)$ and $(u,s)$
Moreover, the tensorial dependence on the polarization vectors
is factored out of the integrals.
The field strength $F^{\mu\nu}$ is the linearized field strength defined by
$F^{\mu\nu} = \varepsilon^{\mu} k^{\nu}-k^{\mu}\varepsilon^{\nu}$
and $R^{\mu\nu\rho\sigma} = F^{\mu\nu}F^{\rho\sigma}$.
The tensor $t_8$ is defined
in \cite[appendix~9.A]{Green:1987mn} in such a way that
$t_8 F^4 = 4 \tr(F^{(1)}F^{(2)}F^{(3)}F^{(4)})
      - \tr(F^{(1)}F^{(2)})\tr(F^{(3)}F^{(4)}) + {\rm perms}\,(2,3,4)$,
where the traces are taken over the Lorentz indices.
In the spinor-helicity formalism, we find
\begin{subequations} \begin{align}
   2{t_8 F^4}& = \spaa 12 ^2 \spbb 34 ^2 \,, \\
   4{t_8 t_8 R^4}& =  \spaa 12 ^4 \spbb 34 ^4 \,.
\end{align} \label{e:prefactors} \end{subequations}

The compactness of the expressions \eqref{e:WX-def}
is characteristic to the worldline formalism.
In particular, the single function $W_X$
determines the whole gauge theory amplitude in all of its kinematic channels.

Note that, contrary to the tree-level case, where integrations by parts
have to be performed to ensure the vanishing of tachyon poles, at one loop,
the field theory limit can be computed without integrating out the double
derivatives.\footnote{At least when there are no triangles.}

\subsection{$\cN=2$ SYM amplitudes from string theory}
\label{sec:st-gauge}

In this section, we provide the string-theoretic integrands for the scattering
amplitudes of four gauge bosons in $\cN=2$ SYM in heterotic string.
Starting from the class of $\cN=2$ four-dimensional heterotic orbifold
compactifications constructed in \cite{Gregori:1998fz,Gregori:1999ns} and
following the recipe of the previous section,
detailed computations have been given
in \cite{Tourkine:2012vx} based on the previously explained method.
We shall not repeat them here but simply state the result.
First of all, we recall that the expansion \eqref{amplitudeN2}
of the $\cN=2$ gluon amplitude into a sum of the $\cN=4$ amplitude
with that of the $\cN=2$ hyper-multiplet.
The corresponding worldline numerators
for the color-ordered amplitudes of eq.~\eqref{e:WX-def} are:
\begin{equation}
      W_{\cN=4,\text{vect}} = 1 \,, ~~~~~~~~~~
      W_{\cN=2,\text{hyper}} = W_3 \,,
\end{equation}
and, according to eq.~\eqref{amplitudeN2}, combine into $W_{\cN=2,\text{vect}}$ as follows:
\begin{equation}
      W_{\cN=2,\text{vect}} = 1 - 2 W_3 \,.
\label{e:Neq2gauge}
\end{equation}
The polynomial $W_3$, derived originally in the symmetric $\cN=4$ supergravity
construction of \cite{Tourkine:2012vx}, is defined by 
\begin{equation}
 W_3 = -{1\over 8}\left(
 (\dot G_{12}-\dot G_{14})
 (\dot G_{21}-\dot G_{24})+
 (\dot G_{32}-\dot G_{34})
 (\dot G_{42}-\dot G_{43})
 \right) \,.
\label{e:W3def}
\end{equation}
where we introduce the shorthand notation $G_{ij} = G(u_i,u_j)$
and, accordingly, $\dot G_{ij}$ are defined in eq.~\eqref{e:Gwl}.
In spinor-helicity formalism, for the gauge choice
\begin{equation}
   (q_1^{\rm ref},q_2^{\rm ref},q_3^{\rm ref},q_4^{\rm ref})
      = (k_3,k_3,k_1,k_1) \,,
\label{e:MHVgauge}
\end{equation}
it writes explicitly as follows:
\begin{equation} \begin{aligned}
      W_3 = & - \frac18\, (\sg(u_1-u_2)+2 u_2-1)(\sg(u_2-u_1)+2 u_1-1) \\
            & + \frac14\, (\sg(u_3-u_2)+2 u_2-1)( u_3- u_2) \,.
\end{aligned} \label{e:W3explicit} \end{equation}

It is of the form $\dot G^2$, so according to the dictionary \eqref{e:dictionary},
it corresponds to four-dimensional box numerators with two powers of loop momentum.
This statement is coherent with the results of the field-theoretic calculation
of section~\ref{sec:ft-loop}, namely,
with the box numerators~\eqref{boxesfinal}.
Moreover, it obviously has no constant term originating from $({\rm sign} (u_{ij}))^2$,
which is consistent with its absence in the loop-momentum expressions.
We also checked that this worldline numerator integrates
to the correct field theory amplitudes~\eqref{N1chiral}.

\subsection{Absence of triangles}
\label{sec:notriangles}

A direct application of the Bern-Kosower formalism immediately rules out the
possibility of having worldline triangles in the field theory limit,
however it is worth recalling the basic procedure to show this.
      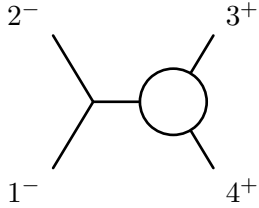
\begin{figure}[t]
      \centering
      \begin{fmffile}{graph21}
      \fmfframe(10,10)(10,10){ \begin{fmfgraph*}(75,50)
            \fmflabel{$1^-$}{g1}
            \fmflabel{$2^-$}{g2}
            \fmflabel{$3^+$}{g3}
            \fmflabel{$4^+$}{g4}
            \fmfleft{g1,g2}
            \fmfright{g4,g3}
            \fmf{plain}{g1,v12,g2}
            \fmf{plain}{g3,v34,g4}
            \fmf{plain}{v12,v34}
            \fmfv{decor.shape=circle,decor.filled=empty,decor.size=0.50h}{v34}
      \end{fmfgraph*} }
      \end{fmffile}
      \caption{$s_{12}$-channel ``would-be'' worldline triangle \label{Striangle}}
      \end{figure}

On the torus, trees attached to loops are produced by vertex operators
colliding to one another, exactly as at tree level.
For instance, consider an $s_{12}$-channel pole,
as drawn in figure~\ref{Striangle}.
It originates from a region of the worldsheet moduli space where $z_{12}\ll1$.
Locally, the worldsheet looks like a sphere, and in particular the short distance
behavior of the torus propagator is as on the sphere:
\begin{equation}
      \cG(z_{12}) = -\frac{\ap}{2} \ln|z_{12}|^2 + O(z_{12}) \,, \qquad
      S_{a,b}(z_{12}) = \frac{1}{z_{12}} + O(z_{12}) \,.
\end{equation}
Repeating the same reasoning as at tree level, a pole will be generated if and only if
a term like $1/|z_{12}|^2$ is present in the numerator factor
$\cW^{\rm (L)}\cZ\overline{ \cW^{\rm (R)}}$.
In the gauge current sector, this requires a term like $S_{a,b}(\bar z_{12})$
that comes along with a single or double trace, like
$\tr(\dots T^{a_1}T^{a_2} \dots)$ or $\tr(T^{a_1}T^{a_2})\tr(T^{a_3}T^{a_4})$,
which causes no trouble. However, in the supersymmetric sector, this term has to be a
$\partial \cG(z_{12})$ which amounts to extraction from $\cW_3$ of the following term:
\begin{equation}
   \partial \cG(z_{12}) (\partial \cG(z_{14}) - \partial \cG(z_{24})) \simeq
      z_{12}\partial \cG(z_{12})\partial ^2 \cG(z_{24}) + O(z_{12})^2 \,,
\end{equation}
which obviously does not provide the expected $1/z_{12}$ behavior.
Note that $\left(\partial \cG(z_{12})\right)^2$ does not work either,
as it is killed by the phase integration.

It is not difficult to check that no triangles are generated
in the other channels, and this is independent of the gauge choice.
As we shall explain later in the comparison section,
our BCJ triangles \eqref{trianglesfinal} are invisible in the worldline formulation,
which is consistent with the previous observation.

We could also try to observe Jacobi identities on $W_3$ directly on the worldline. 
A possible natural way to do so is to consider the following difference:
$ W_3\big|_{u_1<u_2} - W_3\big|_{u_2<u_1}$
and try to associate it to a BCJ triangle.
This quantity, when it is non-zero, can be found to be proportional to $u_i-u_j$.
This definitely vanishes when considering a triangle-like configuration with
coinciding points $u_i \to u_j$.

\subsection{$(2,2)$ $\cN=4$ supergravity amplitudes from string theory}
\label{sec:st-grav}

The four-graviton amplitudes in $(2,2)$ string theory models have been studied
in \cite{Tourkine:2012vx} using type II symmetric orbifold constructions of
\cite{Gregori:1997hi}. 
Here we shall not recall the computation but only describe the structure of the
numerator $\cW^{(L)}\cZ \overline{ \cW^{(R)}}$. 
In the symmetric $(2,2)$ constructions,
both the left-moving and the right-moving sectors of the type II string have
the half-maximal supersymmetry. 
Therefore this leaves room for internal left-right contractions in addition to
the usual chiral correlators when applying Wick's theorem to compute the
conformal blocks. 
Schematically, the integrand can be written as follows:
\begin{equation}
 (1 - 2\cW_3)(1 - 2\overline \cW_3) + 2\cW_2\,,
\end{equation}
where the partition function has explicitly produced a sum over the
orbifold sectors to give $1$ and $-2\cW_3$.

After taking the field theory limit, one obtains the following worldline
numerators for $\cN\!=\!4$ supergravity coupled to two $\cN\!=\!4$ vector
multiplets:
\begin{equation}
      W_{\cN=4,\text{ grav + 2 vect}} = (1 - 2W_3)^2 + 2W_2 \,,
\label{e:Neq4grav}
\end{equation}
where $W_3$ is the same as in eq.~\eqref{e:W3def} and the polynomial $W_2$ writes
\begin{equation}
      W_2 = - (\dot G_{12}-\dot G_{14})(\dot G_{32}-\dot G_{34})\ddot G_{24} \,,
\label{e:W2def}
\end{equation}
in the gauge choice \eqref{e:MHVgauge}, its explicit expression is
\begin{equation}
      W_2 = - {1\over4 T}\;{1\over u}\, (2u_2-1+\sg(u_3-u_2)) 
              (2u_2-1+\sg(u_1-u_2)) \,.
\label{e:W2explicit}
\end{equation}
According to the dictionary \eqref{e:dictionary},
in the field-theoretic interpretation,
$W_3^2$ corresponds to a four-dimensional box numerator
of degree four in the loop momentum,
whereas $W_2$ can be interpreted as a degree-two box numerator in six dimensions,
due to its dimension-shifting factor $1/T$ characteristic of the
left-right-mixed contractions, see eq.~\eqref{e:lr-one-loop-worldline}.
Following the supersymmetry decomposition~\eqref{N4expansion},
we can rewrite eq.~\eqref{e:Neq4grav} as
\begin{equation}
 W_{\cN=4,\text{ grav + 2 vect}} = W_{\cN=8,\text{grav}}-
4W_{\cN=6,\text{matt}}+4W_{\cN=4,\text{matt}}\,,
\end{equation}
where the integrands are given by:
\begin{subequations} \begin{align}
 & W_{\cN=8,\text{grav}} \,= 1 \,,
   \label{e:Neq8}\\
 & W_{\cN=6,\text{matt}} =  W_3 \,,
   \label{e:Neq6matt}\\
 & W_{\cN=4,\text{matt}} =  W_3^2 + W_2/2 \,.
   \label{e:Neq4vm}
\end{align} \label{e:Ngt4} \end{subequations}
These numerators respectively integrate to the following expressions:
\begin{align}
      \cM^{\text{1-loop}}_{\cN=8,\text{grav}} ~ & = \frac{t_8 t_8 R^4}{4} \,
      \bigg\{\frac{2}{\epsilon}
      \bigg[
            \frac{1}{su} \ln\left(-t\over\mu^2\right)
          + \frac{1}{tu} \ln\left(-s\over\mu^2\right)
          + \frac{1}{st} \ln\left(-u\over\mu^2\right)
      \bigg] \label{e:N8} \\
      + 2
      \bigg[
            \frac{1}{st}&\ln\left(-s\over\mu^2\right) \ln\left(-t\over\mu^2\right)
          + \frac{1}{tu} \ln\left(-t\over\mu^2\right) \ln\left(-u\over\mu^2\right)
          + \frac{1}{us} \ln\left(-u\over\mu^2\right) \ln\left(-s\over\mu^2\right)  
      \bigg] \bigg\} \,, \nn\\
      \cM^{\text{1-loop}}_{\cN=6,\text{matt}} & = -\frac{t_8 t_8 R^4}{2 s^2}\,
      \left(\ln^2\left(-t\over-u\right)+\pi^2 \right) \,, \label{e:N3half} \\
      \cM^{\text{1-loop}}_{\cN=4,\text{matt}} & =  \frac{t_8 t_8 R^4}{2 s^4}
      \left[ s^2 + s(t-u)\ln\left(-t\over-u\right)
                 - tu \left( \ln^2\left(-t\over-u\right)+\pi^2 \right)
      \right] \,, \label{e:VM40}
\end{align}
which match to the field theory amplitudes from section~\ref{sec:ft-loop}
($\mu$ being an infrared mass scale).

\section{Comparison of the approaches}
\label{sec:comparison}

In this section, we compare the field-theoretic and the string-based
constructions for gauge theory and gravity amplitudes. We start with the
simplest cases of section~\ref{sec:ft-N4} in which the BCJ construction contains
at least one $\cN=4$ gauge theory copy.

Looking at the string-based representations for $\cN>4$ supergravity amplitudes
in eqs.~\eqref{e:Neq8} and \eqref{e:Neq6matt}, 
one sees that they do verify the double copy prescription, because the
$\cN=4$ Yang Mills numerator $W_{\cN=4,\text{vect}}$ is simply~$1$.
Therefore, regardless of the details of how we interpret the worldline integrand
in terms of the loop momentum, the the double copy prescription~\eqref{e:Mgeneral}
is immediately deduced from the following relations which express the gravity
worldline integrands as products of gauge theory ones:
\begin{subequations} \begin{align}
   W_{\cN=8,\text{grav}} &= W_{\cN=4,\text{vect}} \times  W_{\cN=4,\text{vect}} \,, \\
   W_{\cN=6,\text{matt}} &= W_{\cN=4,\text{vect}} \times  W_{\cN=2,\text{hyper}} \,.
\end{align} \end{subequations}
These $\cN>4$ cases match directly to their field-theoretic construction
described in section~\ref{sec:ft-N4}.
Unfortunately, they do not allow us to say anything about the string-theoretic origin
of kinematic Jacobi identities, as there are no triangles in both approaches,
therefore they require only the trivial identity
     $ 1 - 1 = 0 $.

We can also derive, without referring to the full string-theoretic construction,
the form of the $\cN=6$ supergravity amplitude,
simply by using its supersymmetry decomposition~\eqref{N6expansion}:
\begin{equation}
      W_{\cN=6,\text{grav}} = W_{\cN=4,\text{vect}} \times
           \left( W_{\cN=4,\text{vect}} - 2\,W_{\cN=2,\text{hyper}} \right) \,,
\end{equation}
which, according to eq.~\eqref{e:Neq2gauge}, can be rewritten as
\begin{equation}
      W_{\cN=6,\text{grav}} =  W_{\cN=4,\text{vect}} \times  W_{\cN=2,\text{vect}} \,.
\end{equation}

The first really interesting case at four points
is the symmetric construction of $\cN=4$ gravity with two vector multiplets,
whose string-based numerator was given in eq~\eqref{e:Neq4grav}.
This numerator is almost the square of \eqref{e:Neq2gauge}, up to the term $W_2$
which came from the contractions between left-movers and right-movers.
Due to the supersymmetry expansion~\eqref{N4expansion},
the same holds for the string-based numerator of
$\mathcal{M}_{\cN=4,\text{matt}}^{\text{1-loop}}$.
In the following sections,
we will compare the integrands of that amplitude coming from string and field theory,
and see that the situation is quite subtle.

The aim of the following discussion (and, to a large extent, of the paper)
is to provide a convincing illustration that the presence of total derivatives
imposed by the BCJ representation of gauge theory integrands
in order to obtain the correct gravity integrals
has a simple physical meaning from the point of view of closed string theory. 

As we have already explained,
in the heterotic string construction of Yang-Mills amplitudes,
the left- and right-moving sector do not communicate to each other as
they have different target spaces. 
However, in gravity amplitudes, the two sectors mix due to left-right
contractions.

Our physical observation is that these two aspects are related.
To show this, we will go through a rather technical procedure in order to
compare loop-momentum and Schwinger proper-time expressions,
to finally write the equality~\eqref{e:W2-simp1} of the schematic form
\begin{equation}
	\int \text{left-right contractions} =
      \int \text{(BCJ total derivatives)}^2 + (\,\dots) \,\,.
\label{e:rl-schematic}
\end{equation}

We shall start by the gauge theory analysis and see that,
despite the absence of left-right contractions, the string theory integrand
is not completely blind to the BCJ representation
and has to be corrected so as to match it at the integrand level,
see eq.~\eqref{e:nboxW3}.

On the gravity side, the essential technical difficulty that we will face
is the following: in the two approaches, the squaring is performed in terms of
different variables, and a square of an expression in loop momentum space does
not exactly correspond to a square in the Schwinger proper-time space.
This induces the presence of ``square-correcting terms'',
the terms contained in $(\,\dots)$ on the right-hand side of
eq.~\eqref{e:rl-schematic}.

\subsection{Going from loop momenta to Schwinger proper times}
\label{sec:loop-time}
In principle, there are two ways to to compare loop-momentum expressions to
worldline ones:
one can either transform the loop-momentum into Schwinger proper times, or the
converse.
We faced technical obstacles in doing the latter,
mostly because of the quadratic nature of the gauge theory loop-momentum
polynomials,
so in the present analysis we shall express loop-momentum numerators
in terms of Schwinger proper-time variables.

We use the standard exponentiation procedure
\cite{Bern:1992em,Bern:1993kr}\footnote{See also
\cite{BjerrumBohr:2008vc,BjerrumBohr:2008ji}
for an $n$-point review of the procedure
in connection with the worldline formalism}
which we review here.
First of all, let us consider the scalar box:
\begin{equation}
      I[1] = \int {\d ^d \ld \over (2\pi)^d} {1\over \ld^2
             (\ld-k_1)^2(\ld-k_1-k_2) ^2 (\ld+k_4)^2} \,.
\label{e:scalarbox}
\end{equation}
We exponentiate the four propagators using
\begin{equation}
      \frac{1}{D_i^2} = \int_0^\infty\d x_i \exp(-x_i D_i^2) \,,
\label{e:stand-exp}
\end{equation}
and obtain
\begin{equation}
   I[1] = \!\!\!\int\!\frac{\d^d \ld}{(2\pi)^d~}
            \!\!\int_0^\infty\!\!\prod_{i=1}^4 \d x_i
          \exp\!\left(\! - \ld^2\!\sum_{i=1}^4 x_i
                   \!+\! 2 \ld\!\cdot\!(x_2 k_1 \!+\! x_3 (k_1\!+\!k_2) \!-\! x_4 k_4)
                   \!-\! x_3 (k_1\!+\!k_2)^2 \!\right) \,,
\end{equation}
after expanding the squares. 
Then we introduce the loop proper time $T=\sum_i x_i$ and rescale the $x_i$'s
by defining the standard Feynman parameters
\begin{equation}
      a_i = \frac{x_i}{T} \,,
\label{e:ai}
\end{equation}
which gives:
\begin{equation}
      I[1] =\int_0^\infty \!\! {\d T}\,T^3 \int {\d ^d \ld\over (2\pi)^d} \int_0^1
      \prod_{i=1}^4 \d a_i\,\delta\left(1-\sum_{i=1}^4 a_i\right)
      \exp\left( -T (\ld-K)^2 -T Q \right) \,.
\end{equation}
In this expression, the scalar $Q$ is the second Symanzik polynomial and is
given by
\begin{equation}
      Q = - a_1 a_3 s - a_2 a_4 t \,,
\label{e:Q}
\end{equation}
while the shift vector $K=(x_2 k_1 + x_3 (k_1+k_2) - x_4 k_4)$
defines the shifted loop momentum
\begin{equation}
      \tld=\ld-K\,.
\label{e:ltilde}
\end{equation}
Of course, the expressions for $Q$ and $K$ change
with the ordering in this parametrization.

\begin{figure}[t]
      \centering
      \includegraphics[scale=1]{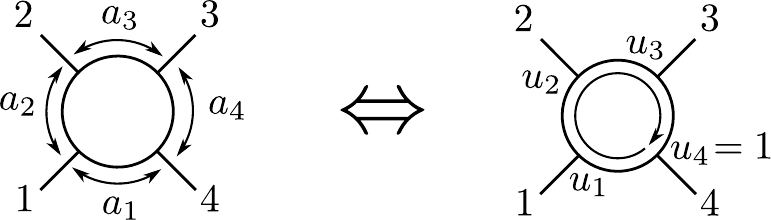}
\caption{The $u_i$'s are unambiguously defined for all orderings.}
\label{fig:WLua}
\end{figure}

If we go to the worldline proper times $t_i$,
or rather their rescaled versions
\begin{equation}
      u_i = \frac{t_i}{T} \,,
\label{e:ui}
\end{equation}
defined as sums of the Feynman parameters, as pictured in figure~\ref{fig:WLua},
one obtains a parametrization valid for any ordering of the legs,
in which the vector $K$ writes \cite{BjerrumBohr:2008ji,BjerrumBohr:2008vc}
\begin{equation}
      K^\mu = -\sum u_i {k_i}^\mu \,.
\end{equation}
The scalar $Q$ also has an invariant form in these worldline parameters,
already given in~\eqref{e:Qtrop}.
Finally, the Gaussian integral in $\tld$ is straightforward to perform, and we
are left with:
\begin{equation}
      I[1] = \frac{i}{(4\pi)^{d/2}}\int_0^\infty {\d T\over T^{d/2-3}}
             \int \prod_{i=1}^3 \d u_i\,
             \exp\left( -TQ \right) \,.
\label{e:scalarbox2}
\end{equation}
In \eqref{e:scalarbox2}, the integration domain $\{0<u_1<u_2<u_3<1\}$, gives
the box \eqref{e:scalarbox} ordered as $(k_1,k_2,k_3,k_4)$,
whereas the two other orderings are given by the integration domains
$\{0<u_2<u_3<u_1<1\}$ and $\{0<u_3<u_1<u_2<1\}$.

\subsection{Comparison of gauge theory integrands}
\label{sec:comp-gauge}

Now we can repeat the same procedure for a box integral $I[n(\ell)]$
with a non-trivial numerator. 
Our BCJ box numerators \eqref{boxesfinal} are quadratic
in the four-dimensional loop momentum\footnote{We recall that the $d$- and four-
dimensional loop momenta are related by $\ld^2 = \ell ^2-\mu^2$.} and can be
schematically written as:
\begin{equation}
   n^{(\mathfrak{S})}_{\rm box} (\ell)
      = A_{\mu \nu}\ell^\mu\ell^\nu + B^{(\mathfrak{S})}_\mu \ell^\mu \,,
\label{e:nbox}
\end{equation}
where the label $\mathfrak{S}$ refers to one of the inequivalent orderings
$\{(123),(231),(312)\}$.
One can verify that the quadratic form $A_{\mu\nu}$ does not depend on the ordering.
Note that we did not write the constant term in eq.~\eqref{e:nbox},
because there are none in our master BCJ boxes \eqref{boxesfinal}.
The exponentiation produces an expression which depends
both on Schwinger proper times and the shifted loop momentum:
\begin{equation}
   n^{(\mathfrak{S})}_{\rm box}(\tell+K) = A_{\mu \nu}\tell^\mu\tell^\nu
                   + (2 A_{\mu\nu} K^\nu + B^{(\mathfrak{S})}_\mu) \tell^\mu
                      + A_{\mu\nu} K^\mu K^\nu + B^{(\mathfrak{S})}_\mu K^\mu \,.
\end{equation}
The linear term in $\tell$ integrates to zero in the gauge theory amplitude,
but produces a non-vanishing contribution when squared in the gravity
amplitude.
Lorentz invariance projects $A_{\mu \nu}\tell^\mu\tell^\nu$ on its trace,
which turns out to vanish in our ansatz:
\begin{equation}
      \tr A = 0 \,.
\label{e:trA=0}
\end{equation}
Then we define $\langle n^{(\mathfrak{S})}_{\rm box} \rangle$ to be
the result of the Gaussian integration over $\tell$:
\begin{equation}
      \langle n^{(\mathfrak{S})}_{\rm box} \rangle = A_{\mu\nu}K^\mu K^\nu
            + B^{(\mathfrak{S})}_\mu K^\mu \,.
\label{e:nboxbracket}
\end{equation}
Note that here and below, for definiteness and normalization,
we use the bracket notation $\langle...\rangle$ for integrand numerators
in terms of the rescaled Schwinger proper times $u_i$
so that $I[n]$ can be written in any integration-parameter space:
\begin{equation}
\begin{aligned}
      \int\!\!\frac{\d^d \ell}{(2\pi)^d}
            \frac{n(\ell)}{\ld^2 (\ld-k_1)^2 \dots (\ld-\sum_{i=1}^{n-1}k_i)^2}
          = \frac{(-1)^n i}{(4\pi)^{\frac{d}{2}}}
            \int_0^\infty\!\!\!{\d T\over T^{\frac{d}{2}-(n-1)}}\!
            \int \prod_{i=1}^{n-1} \d u_i \braket{n} e^{-T Q} \,,
\label{e:bracket}
\end{aligned}
\end{equation}
where the integration domain in $u_i$ corresponds to the momentum ordering
in the denominator.
From the previous reasoning, it is easy to show the following dictionary:
a polynomial $n(\ell)$ of degree $k$ in the loop momentum
is converted to a polynomial $\braket{n}$ of the same degree in Schwinger proper times: 
\begin{equation}
      n(\ell) = O(\ell^k) \quad \Rightarrow \quad \braket{n} =
	O\left(\frac{u^{k-2p}}{T^p}\right) \,,\quad p\geq0
\label{e:bracketdict}
\end{equation}
where the inverse powers of $T^p$ correspond to terms of the form ${\tell}^{2p}$
and both consistently act as dimension shifts, as it can be seen on the
standard replacement rules given later in eq.~\eqref{e:lreplacement}. This is
consistent with
\eqref{e:dictionary}.  
We can recast the previous
procedure in table~\ref{tab:comparison}, to summarize the common
points between the worldline formalism and usual Feynman diagrams.
\begin{table}[h]
\centering
\begin{tabular}{|c|c|c|}
 \hline
  &
  \hspace{1pt} Field theory \hspace{1pt} &
  \hspace{1pt} Worldline \hspace{1pt} \\
  \hline
  \hspace{1pt} Parameters  \hspace{1pt} &
  \hspace{1pt} $\ell_{(d)}$ \hspace{1pt} &
  \hspace{1pt} $u_i, T$ \hspace{1pt} \\
  \hspace{1pt} Numerator \hspace{1pt} &
  \hspace{1pt} $n(\ell)$ \hspace{1pt} &
  \hspace{1pt} $W_X,\langle n \rangle$ \hspace{1pt} \\
  \hspace{1pt} Denominator \hspace{1pt} &
  \hspace{1pt} $\tfrac{1}{D(\ell)}$ \hspace{1pt} &
  \hspace{1pt} $e^{-TQ}$ \hspace{1pt} \\
  \hspace{1pt} Power counting \hspace{1pt} &
  \hspace{1pt} $O(\ell^k)$ \hspace{1pt} &
  \hspace{1pt} $O(u_i^k)$ \hspace{1pt} \\
 \hline
\end{tabular}
\caption{Basic ingredients of the loop integrand expressions in field theory
and the field theory limit of string theory.}
\label{tab:comparison}
\end{table}

We apply this method  to the BCJ box numerators,
in order to compare them to the string-based numerator $W_3$.
These two quantities have equal integrated contributions, as was noted before.
However, at the integrand level, they turn out not to be equal.
We denote their difference by $\delta W_3$:
\begin{equation}
   \langle n^{(\mathfrak{S})}_{\rm box} \rangle = W_3 + \delta W_3 \,.
\label{e:nboxW3}
\end{equation}
By definition, $\delta W_3$ integrates to zero separately in each sector.

Making contact with the tree-level analysis, where the integrands had to be
put in the particular MSS representation in string theory to ensure the manifest BCJ
duality, one can wonder if this term $\delta W_3$ has a similar meaning at
one loop.
We note that the information that it carries, of order $\ell^2$,
is not trivial and is sensitive to the BCJ solution,
since the quadratic terms in the box numerators \eqref{boxesfinal} are fixed
to comply with the kinematic Jacobi identities.

Therefore, $\delta W_3$ seems to be a messenger of the BCJ representation
information and indicate a specific worldline representation of the string
integrand.
In order to be more precise on this statement, let us first rewrite
$\delta W_3$ in terms of worldline quantities, i.e. as a polynomial in the
worldline Green's functions.
As it is of order $u_i^2$, it has to come from a polynomial with at most
binomials of the form $\dot G_{ij}\dot G_{kl}$.
By a brute-force ansatz, we have expressed $\delta W_3$
as a function of all possible quantities of that sort.
Imposing the defining relation \eqref{e:nboxW3} in the three sectors results in 
{a three-parameter space of possibilities for $\delta W_3$}
(see the full expression~\eqref{e:deltaWgen} in the appendix).
All consistency checks were performed on this numerator. 
At this level, the parameters $\alpha$ and $\beta$ of the BCJ numerators
\eqref{boxesfinal}, \eqref{trianglesfinal} are still free. 
It turns out that they create a non-locality in the worldline integrand, of
the form $tu/s^2$. To cancel it, one has to enforce the condition
\begin{equation}
   1-\alpha+\beta=0\,,
\label{e:local-WL}
\end{equation}
consistent with the choice \eqref{e:A7A11conj}.
Below we provide a representative element of the family of $\delta W_3$'s
that we obtained from our ansatz:
\begin{equation} \begin{aligned} \label{e:deltaWex}
   \delta W_3 = \frac{1}{2}
      \big( 2 \dot G_{12}^2 - 2 \dot G_{34}^2
 	      -(\dot G_{23}\!-\!3\dot G_{14})(\dot G_{13}\!-\!\dot G_{24})
          & + \dot G_{12} (\dot G_{23} - \dot G_{13} + 3\dot G_{24} - 3\dot G_{14}) \\
          & + \dot G_{34} (3\dot G_{14} - 3\dot G_{13} - \dot G_{23} + \dot G_{24})
      \big)\,.
\end{aligned} \end{equation}

In order to safely interpret $\delta W_3$ as a natural string-based object,
it is important to verify that its string ancestor would not create any
triangles in the field theory limit.
We will refer to this condition as the
``\emph{string-ancestor-gives-no-triangles}'' criterion. 
This is not a trivial property, and it can be used to rule out some
terms as possible worldline objects
(see, for example, the discussion in appendix~\ref{app:sqt}).
In the present case, it was explicitly checked that the full form of $\delta
W_3$ given in appendix~\ref{app:dW} satisfies this property,
following the procedure recalled in section~\ref{sec:notriangles}.

Now that we have expressed $\delta W_3$ in this way, let us look back at what
is the essence of the tree-level MSS approach. It is based on the fact that the
correct tree-level form of the integrand is reached after a series of
integration by parts.\footnote{In sec.~\ref{sec:st-tree} we did
not have to perform any due to a sufficiently restrictive gauge
choice.}
One might hope that the worldline numerator defined by $W_3+\delta W_3$
is actually a result of application of a chain of integration by parts.
Unfortunately, we have not found any sensible way in which the worldline
numerator $W_3+\delta W_3$ could be obtained from $W_3$ by such a process.
The reason for this is the presence of squares in $\delta W_3$,
of the form $\dot G_{ij}^2$, which are not possible to eliminate
by adjusting the free parameters of eq.~\eqref{e:deltaWgen}.
These terms are problematic for basically the same reason as at tree level,
where, to integrate them out by parts,
you always need a double derivative and a double pole to combine together.
This can be seen at one loop by inspecting the following identity:
\begin{equation}
      \partial_1 \left(\dot G_{12} e^{-T Q}\right) 
    = \ddot G_{12} + 
      \dot G_{12} \left( k_1\!\cdot k_2\,\dot G_{12}
                       + k_1\!\cdot k_3\,\dot G_{13}
                       + k_1\!\cdot k_4\,\dot G_{14} \right) \,,
\end{equation}
where we see that the square $\dot G_{12}^2$ always goes {in pair}
with the double derivative $\ddot G_{12}$.
A similar equation with $\partial_1$ replaced with $\partial_2$ does not help,
as the relative signs between the double derivative and the square are unchanged.
This kind of identities show that, in the absence of double derivatives in
$\delta W_3$, $W_3$ and $(W_3+\delta W_3)$ are not related by a chain of
integration by parts.
The reason why we cannot include these double derivatives in our ansatz for
$\delta W_3$ is
because they would show up as $1/T$ terms in eq.~\eqref{e:nboxbracket} which is
impossible in view of the tracelessness of $A_{\mu\nu}$,
eq.\eqref{e:trA=0}.

Therefore, the introduction of $\delta W_3$ in the string integrand to make it
change representation, although not changing the integrated result and
satisfying this ``string-ancestor-gives-no-triangles'' property, appears to be a
non-IBP process, in contrast with the MSS procedure. 
It would be interesting to understand if this property is just an artifact
of our setup, or if it is more generally a sign that string theory does not
obey the full generalized gauge invariance of the BCJ representation.

Finally, we note that $\delta W_3$ is not directly related to the BCJ triangles.
Recall that they are defined through the BCJ color-kinematics duality
and are crucial for the double copy construction of gravity.
But in section~\ref{sec:notriangles}, we saw that there are no triangles
in our string-theoretic construction.
So even though we find total derivatives both on the field theory
side:\footnote{In eq.~\eqref{e:nboxtri},
we omitted the denominators for notational ease.}
\begin{equation}
      \sum n_{\rm box} + \sum n_{\rm tri} ,
\label{e:nboxtri}
\end{equation}
and on the worldline side in the BCJ inspired form:
\begin{equation}
      W_3 + \delta W_3 \,,
\label{e:nboxdW}
\end{equation}
where the BCJ triangles and $\delta W_3$ integrate to zero,
they cannot be made equal by going to proper-time variables, as\footnote{See
appendix~\ref{app:triangles} for more details on eq.~\eqref{e:triangleWL}.}
\begin{equation}
      \braket{n_{\text{tri}}} = 0 \,.
\label{e:triangleWL}
\end{equation}
In addition to that, $\delta W_3$ is truly as a box integrand.

In any case, the important point for the next section is that both $\delta W_3$
and
the BCJ triangles contribute to the gravity amplitude when squared.
We will try to relate them to the new term $W_2$ that appears in gravity
from left-right mixing terms.

\subsection{Comparison of gravity integrands}
\label{sec:comp-grav}

The goal of this final section is to dissect the BCJ gravity numerators
obtained by squaring the gauge theory ones in order to perform a thorough comparison
with the string-based result. 
In particular, we wish to illustrate
that the role of the left-right contractions is to provide the terms
corresponding to the squares of the total derivatives in the loop momentum space
(the BCJ triangles and the parity-odd terms).

\subsubsection{String-based BCJ representation of the gravity integrand}

At the level of integrals, we can schematically equate the gravity amplitude
obtained from the two approaches:
\begin{equation}
	\int \sum \langle n_{\rm box}^2 \rangle
         + \sum \langle n_{\rm tri}^2 \rangle = \int W_3^2 + W_2/2 \,,
\end{equation}
where we omitted the integration measures and the factors of $\exp({-TQ})$.
In order to relate the left-right contractions in $W_2$
to the triangles $\braket{n_{\rm tri}^2}$,
we first need to consider the relationship between the squares $W_3^2$
and $\langle n_{\rm box}^2\rangle$ via $\langle n_{\rm box}\rangle^2$, using
the result of the previous section.
From eq.~\eqref{e:nboxW3}, we know that at the gauge theory level,
the integrands match up to a total derivative $\delta W_3$.
Therefore, let us introduce by hand this term in the string-based gravity integrand:
\begin{equation}
	 W_3^2 +  W_2/2 =
	\langle n_{\rm box}\rangle^2 +  W_2/2 - (2 W_3 \delta W_3 + {\delta
W_3}^2)\,.
\end{equation}
The cost for this non-IBP change of parametrization is the introduction
of a correction to $W_2$, that we call $\delta W_2$, in the string-based
integrand:
\begin{equation}
      \delta W_2 = -2(2 W_3 \delta W_3 + \delta W_3^2) \,.
      \label{e:deltaW2}
\end{equation}
Note that this term \emph{is not} a total derivative.
%
%
The meaning of this correcting term is that,
when we change $W_3$ to $\braket{n_{\rm box}}$, we also have to modify $W_2$.
In this sense, it is induced by the Jacobi relations
in the gauge theory numerators $n_{\rm box}$. 
Moreover, had we managed to do
only integration by parts on $W_3^2$,
$W_2$ would have received corrections due to the left-right contractions
appearing in the process. These would show up as factors of $1/T$, as already
explained below eq.~\eqref{e:lr-one-loop-worldline}.

%

Again, to be complete in the interpretation of $\delta W_2$ as a proper
worldline
object, we should make sure that it obeys
the ``string-ancestor-gives-no-triangles'' criterion, as we did for $\delta W_3$.
Since we have a symmetric construction for the gravity amplitude, it is natural
to assume that both sectors would contribute equally to this string-theoretic
correction: 
\begin{equation}
      \delta \cW_2=-2 \left(\cW_3 \overline{\delta \cW_3}
                    + \overline{\cW_3} {\delta \cW}_3 + |\delta
			\cW_3|^2\right)\,.
\label{e:cW20}
\end{equation}
Following the analysis of section~\ref{sec:notriangles},
it is easy to convince oneself that since neither $\cW_3$ nor $\delta \cW_3$
gave any triangles in gauge theory, any combination thereof will not either.

Therefore, it seems legitimate to interpret $\delta W_2$ as a string-based
correction,
and this lets us rewrite the worldline numerator of the gravity amplitude as
\begin{equation} \begin{aligned}
      \int \sum \langle n_{\rm box}^2 \rangle
         + \sum \langle n_{\rm tri}^2 \rangle  =
      \int \sum \langle n_{\rm box} \rangle^2 + (W_2+\delta W_2)/2 \,.
\label{e:N4matterdeltanum}
\end{aligned} \end{equation}

\subsubsection{Loop momentum squares vs. worldline squares}

The next step is to relate $\braket{n^2_{\rm box}}$ to $\braket{n_{\rm box}}^2$.
Let us first look at the gravity box numerator.
As before, it can be written as a function of the shifted loop momentum $\tell$:
\begin{equation} \begin{aligned}
      \left( n_{\rm box}(\tell+K) \right)^2
            = & \tell^\mu \tell^\nu \tell^\rho \tell^\sigma
                A_{\mu\nu} A_{\rho\sigma} \\
            + & \tell^\mu \tell^\nu
                \left((2A_{\mu\rho}K^\rho + B_\mu) (2A_{\nu\sigma}K^\sigma + B_\nu)
            + 2 A_{\mu\nu} (A_{\rho\sigma}K^\rho K^\sigma + B_\rho K^\rho) \right)\\
            + & (A_{\rho\sigma}K^\rho K^\sigma+B_\rho K^\rho)^2 \,,
\end{aligned} \end{equation}
where we omitted the terms odd in $\tell$ since they integrate to zero.
Notice, however, that the terms of $n_{\rm box}$ linear in $\tell$,
which used to be total derivatives in gauge theory, now contribute to the
integral, with the $\epsilon(k_1,k_2,k_3,\ell)^2$ terms inside them.
To obtain the proper-time integrand $\langle n_{\rm box}^2 \rangle$,
we go again through the exponentiation procedure of section~\ref{sec:loop-time},
followed by a dimension shift \cite{Bern:1995db},
together with the standard tensor reduction:\footnote{Remember
that the numerator loop momentum $\tell$ is strictly four-dimensional
and integration is over the $d$-dimensional $\tld$.}
\begin{subequations} \begin{align}
   \tell^\mu \tell^\nu & \rightarrow - {\eta^{\mu\nu}\over 2T} \,, \\
   \tell^\mu \tell^\nu \tell^\rho\tell^\sigma & \rightarrow
            {\eta^{\mu(\nu}\eta^{\rho\sigma)} \over 4T^2} \,,
\end{align} \label{e:lreplacement} \end{subequations}
where
$\eta^{\mu(\nu}\eta^{\rho\sigma)}$ stands for $\eta^{\mu\nu}\eta^{\rho\sigma}
+ \eta^{\mu\rho} \eta^{\nu\sigma} + \eta^{\mu\sigma}\eta^{\nu\rho}$.
We obtain:
\begin{equation}
      \langle n_{\rm box}^2 \rangle =
  {\eta^{\mu(\nu}\eta^{\rho\sigma)} A_{\mu\nu}A_{\rho\sigma}\over 4 T^2}
  -{(2A_{\mu\nu}K^\nu+B_\mu)^2\over 2T}
  +(A_{\rho\sigma}K^\rho K^\sigma+B_\rho K^\rho)^2 \,,
\end{equation}
or, equivalently, using  \eqref{e:nboxbracket},
\begin{equation}
   \langle n_{\rm box}^2 \rangle - \langle n_{\rm box} \rangle^2
      = \frac{\eta^{\mu(\nu} \eta^{\rho\sigma)} A_{\mu\nu} A_{\rho\sigma}}{4T^2}
      - \frac{(2A_{\mu\nu} K^\nu + B_\mu)^2}{2T} \,.
\label{e:nboxsqt}
\end{equation}
This formula describes precisely how squaring in loop momentum space
is different from squaring in the Schwinger parameter space,
so we will call the terms on the right-hand side of \eqref{e:nboxsqt}
\emph{square-correcting terms}.
Note that the fact that there are only $1/T^k$ with $k>0$ terms on the
right-hand side of eq.~\eqref{e:nboxsqt} is not accidental, and would have hold
even without the tracelessness of $A$, eq.~\eqref{e:trA=0}. It can indeed be
seen in
full generality that squaring and the bracket operation do commute at the level
of the $O(T^0)$ terms, while they do not commute at the level of the $1/T^k$.
Below we connect this with the structural fact that left-right contractions
naturally yield $1/T$ terms.
In appendix~\ref{app:sqt}, we also provide another description of these terms
based on a trick which lets us rewrite the $1/T^2$ terms as worldline quantities.

\subsubsection{Final comparison}
\label{sec:final}

Using eq.~\eqref{e:N4matterdeltanum}, we rewrite the contribution of
$W_2+\delta W_2$ at the integrated level as follows:
\begin{equation} \boxed{
	\int \frac{1}{2} (W_2+\delta W_2) 
	=\int \sum \langle n_{\rm tri}^2 \rangle
	 +\frac{\eta^{\mu(\nu} \eta^{\rho\sigma)} A_{\mu\nu}
	A_{\rho\sigma}}{4T^2}
      - \sum_{\mathfrak{S}} \frac{(2A_{\mu\nu} K^\nu + B^{(\mathfrak{S})}_\mu)^2}{2T}
}\label{e:W2-simp1}
\end{equation}

In total, we have argued that the total contribution on the left-hand side is a
modification of $W_2$ generated by the BCJ representation of the gauge theory numerators
in a non-IBP-induced way.
This was supported by the aforementioned ``string-ancestor-gives-no-triangles''
criterion satisfied by $\delta W_2$.
We are now able to state the conclusive remarks on our interpretation
of eq.~\eqref{e:W2-simp1}. Its right-hand side is made of two parts,
of different physical origin:
\begin{itemize}
	\item[--] the squares of gauge theory BCJ triangles,
	\item[--] the square-correcting terms.
\end{itemize}
Note that some of the latter come from the contributions of the gauge theory integrand
which were linear in the loop momentum, including the parity-odd terms $ \epsilon_{\mu\nu\rho\sigma} k_1^\nu k_2^\rho k_3^\sigma $ present in $B^{(\mathfrak{S})}$.

Formula~\eqref{e:W2-simp1} shows clearly the main observation of this paper:
the squares of the total derivatives introduced into the gravity amplitude
by the BCJ double copy construction physically come
from the contractions between the left- and right-moving sectors in string theory.
At a more technical level, the contribution of these contractions
to the string-based integrand also had to be modified to take into account for
the BCJ representation of the gauge theory amplitudes.

This being said, the presence of the square-correcting terms on the right-hand
side deserves a comment.
They contain the dimension-shifting factors of $1/T$, characteristic of the
left-right contractions, as already mentioned.
It is therefore not surprising, that the square-correcting terms show
up
on the right-hand side of eq.~\eqref{e:W2-simp1}, since the left-hand side
is the (BCJ modified) contribution of the left-right contractions.

More interestingly, this seems to suggest that it should be possible to absorb
them into the left-right mixing terms systematically by doing IBP's at the
string theory level.
However, if one considers the worldline polynomials corresponding to $(2AK+B)/T$,
they imply a string-theoretic ancestor of the form
$\partial \bar \partial \cG \times \sum \partial \cG \overline{\partial \cG}$
which eventually does not satisfy the ``string-ancestor-gives-no-triangles''
criterion.\footnote{We have checked that the arising triangles are not the same as our
BCJ triangles squared.}
Therefore, not all of the square-correcting terms possess
a nice worldline interpretation, and this makes the situation not as simple as
doing IBP's. 
This fact is to be connected with the impossibility to obtain the BCJ worldline
gauge theory numerator $W_3+\delta W_3$ by integrating by parts $W_3$ in our
setup.
 
Perhaps the main obstacle here is that the vanishing of the BCJ triangles after
integration does not exactly correspond to the absence of string-based triangles before
integration.
All of this suggests that there might exist BCJ representations which cannot be
obtained just by doing integrations by parts. The characterization thereof in
terms of the subset of the generalized gauge invariance respected by string theory 
would be very interesting.
For instance, it might be that our choice to put all the BCJ bubbles to zero,
allowed by the generalized gauge invariance, is not sensible
in string theory for this particular amplitude with the gauge choice
\eqref{e:MHVgauge}.

Notwithstanding, we believe that the main observations of our paper concerning
the observation that the BCJ representation \textit{can be seen} in string
theory and the physical origin of the squares of total derivatives and
observation that the BCJ construction is related to the presence
of left-right mixing terms in string theory holds very generally.

\section{Discussion and outlook}

In this paper, we have studied various aspects of the BCJ double copy
construction. At tree level, we used the MSS chiral block representation
both in heterotic and type II closed strings to rewrite the five-point field
theory amplitudes in a way in which color factors can be freely interchanged
with kinematic numerators to describe scattering amplitudes of cubic color
scalars, gluons or gravitons.
In this context, the Jacobi identities of~\cite{Mafra:2011kj} appear
as consequences of the MSS representation and are on the same footing as the
equivalence between color and kinematics. In particular, we did not have to use
them to write down the final answer.
Working out the $n$-point combinatorics in the lines of our five-point
example would constitute a new direct proof of the color-kinematics duality at
tree level.

At one loop, we performed a detailed analysis of four-point amplitudes in
$\cN\!=\!4$ supergravity from the double copy of two $\cN\!=\!2$ SYM theories,
both in field theory and the worldline limit of string theory.
This symmetric construction automatically requires adding two matter vectors
multiplets to the gravity spectrum.
Our choice of the BCJ ansatz for which the BCJ bubbles were all set to zero is an
effective restriction of the full generalized gauge invariance of the BCJ
numerators.
We focused on the non-trivial loop-momentum structure of
the BCJ gauge theory integrands, which we expressed as worldline quantities to
make comparison with the string-based ones.

The major drawback of this procedure is that, in the process,
one loses some of the information contained in the loop-momentum gauge theory
numerators.
For example, our BCJ gauge theory triangles turned out to vanish after
integration in this
procedure, so one could think that they are invisible for string theory.
However, the box numerators match the string-based numerator up to a
new term that we called $\delta W_3$. This term integrates to zero in each
kinematic channel, thus guaranteeing matching between the two approaches.
This total derivative $\delta W_3$ shifts the string-based integrand
to the new representation $W_3+\delta W_3$.
We argued that this process is not IBP-induced, in the sense that
$W_3+\delta W_3$ cannot be obtained simply by integrating $W_3$ by parts.
We gave a possible clue to interpret this puzzle, based on the fact that the
restriction of the generalized gauge invariance might be incompatible with
string theory in the gauge choice~\eqref{e:MHVgauge}.
It would be interesting to investigate this point further.

At the gravity level, we wanted to argue that the characteristic ingredients of
the BCJ double copy procedure, namely the squares of terms required by the kinematic
Jacobi identities, are generated in string theory by the left-right
contractions.

The first observation is that, going to the non-IBP-induced representation
$W_3 \to W_3 + \delta W_3$ in the string-based integrand
induces a modification of the left-right mixing terms, $W_2 \to W_2 + \delta W_2$,
which can be safely interpreted as a worldline quantity,
because it obeys the ``string-ancestor-gives-no-triangles'' criterion.

Furthermore, the difference between squaring in loop momentum space and in
Schwinger proper time space induces the square-correcting terms. 
We related them to $W_2+\delta W_2$ and observed that they are of the same
nature as the left-right mixing terms in string theory. Such terms are generically
obtained from IBP's, which suggests that the right process (if it exists) in
string theory to recover full BCJ construction makes use of worldsheet integration
by part, just like in the MSS construction at tree level. However, these square-correcting terms do not obey the ``string-ancestor-gives-no-triangles'' property, which
makes them ill-defined from the string-theoretic point of view.
We suppose that the issues of the non-IBP nature of $\delta W_2$ and
$\delta W_3$ might come from the incompatibility between our restriction
of generalized gauge invariance and our string-based
computation in the gauge \eqref{e:MHVgauge}. 

In any case, this shows that string theory has something to say about
generalized gauge invariance. We believe that this opens very interesting questions
related to the process of finding BCJ numerators by the ansatz approach and to
a possible origin of this generalized gauge invariance in string theory.

Finally, we present the bottom line of our paper in the formula~\eqref{e:W2-simp1}: 
we identified a representation of the left-right mixing terms in which they are
related to the squares of the BCJ triangles and the squares of
parity-odd terms $(i\epsilon_{\mu\nu\rho\sigma}k_1 ^\mu k_2^\nu k_3^\rho
\ell^\sigma)^2$. Besides the previous discussion on the nature of the
square-correcting terms in the right-hand side of eq.~\eqref{e:W2-simp1},
we believe this sheds some light on the a priori surprising fact
that total derivatives in the BCJ representation of gauge
theory amplitudes play such an important role.
The physical reason is deeply related to the structure of the closed string:
in the heterotic string, the left-moving sector does not
communicate with the right-moving one in gluon amplitudes,
while this happens naturally in gravity amplitudes in the type II string
and generates new terms, invisible from the gauge theory perspective.

Concerning further developments, in addition to the open issues that we already
mentioned, it would be natural to explore the possibility that the MSS chiral
blocks might be generalized to loop level amplitudes and understand the role of
$\delta W_3$ and generalized gauge invariance in this context.
For that, one would have to account for the left-right mixing terms, generated
naturally by worldsheet integration by parts, which must play a central role
starting from one loop.
Such an approach, if it exists, would amount to disentangling the two sectors
in a certain sense, and a string-theoretic understanding of such a procedure
would be definitely very interesting.

\section*{Acknowledgements}

It is a pleasure to acknowledge many interesting discussions over the
last months with Zvi Bern, Ruth Britto, John Joseph Carrasco, Henrik Johansson,
Gregory Korchemsky, Carlos Mafra, Oliver Schlotterer, Stephan Stieberger and
Pierre Vanhove.
One of the authors, PT, would like to thank in particular Oliver Schlotterer
for sharing notes at an early stage of the project and discussions about the
role of the parity-odd terms squared in $\cN\!=\!4$ SYM at $n\geq5$ points and their
relationship with left-right mixing terms in closed string theory.
We are also grateful to Ruth Britto, Pierre Vanhove and in particular Oliver
Schlotterer for their helpful comments on the manuscript.
PT would like to thank DAMTP and the AEI for hospitality at various stages of
the project.

This research is partially supported by the ANR grant 12-BS05-003-01, the ERC
Advanced grant No. 247252 and the CNRS grant PICS 6076.

\appendix

\section{Integrals}
\label{app:integrals}

      We adopt the conventional definition
\cite{Bern:1992em,Bern:1993kr}
for dimensionally-regularized massless scalar integrals:
\begin{equation}
      I_n^d = (-1)^{n+1}(4\pi)^{\frac{d}{2}}i
              \int \frac{\d^d \ell_{(d)}}{(2\pi)^d}
              \frac{ 1 }
                   { {\ell_{(d)}}^2 (\ell_{(d)}-k_1)^2 \dots(\ell_{(d)}+k_n)^2 } \,,
\label{integrals}
\end{equation}
where by default $d = 4-2\epsilon$.
Here we give only the integrals, relevant for this paper,
i.e. the zero-mass box:
\begin{equation}
      I_4(s,t) = \frac{2 r_{\Gamma}}{st}
                 \bigg\{
                 \frac{1}{\epsilon^2}
                       \left( (-s)^{-\epsilon} + (-t)^{-\epsilon} \right)
                     - \frac{1}{2} \left( \ln^2 \left(\frac{-s}{-t}\right)
                                        + \pi^2 \right)
                 \bigg\}
               + O(\epsilon) \,,
\label{box0m4d}
\end{equation}
and the one-mass triangle:
\begin{equation}
      I_3(t) = \frac{r_{\Gamma}}{\epsilon^2}
                 \frac{(-t)^{-\epsilon}\!\!\!}{(-t)} \,,
\label{tri1m4d}
\end{equation}
where
\begin{equation}
      r_{\Gamma} = \frac{ \Gamma(1+\epsilon) \Gamma^2(1-\epsilon) }
                        { \Gamma(1-2\epsilon) } \,.
\label{rgamma}
\end{equation}

      We also encounter six-dimensional versions of these integrals:
\begin{equation}
      I_4^{d=6-2\epsilon}(s,t) = - \frac{r_{\Gamma}}{2(1-2\epsilon)(s+t)}
                                   \left( \ln^2 \left(\frac{-s}{-t}\right)
                                        + \pi^2 \right)
                                 + O(\epsilon) \,,
\label{box0m6d}
\end{equation}
\begin{equation}
      I_3^{d=6-2\epsilon}(t)
      = \frac{r_{\Gamma}}{2 \epsilon (1-\epsilon) (1-2\epsilon)}
                         (-t)^{-\epsilon} \,,
\label{tri1m6d}
\end{equation}
as well as the eight-dimensional box:
\begin{equation}
      I_4^{d=8-2\epsilon}(s,t) =   \frac{r_{\Gamma}}{(3-2\epsilon)(s+t)}
                                   \left( \frac{st}{2} I_4^{d=6-2\epsilon}(s,t)
                                        + s I_3^{d=6-2\epsilon}(s)
                                        + t I_3^{d=6-2\epsilon}(t)
                                   \right)
                                 + O(\epsilon) \,.
\label{box0m8d}
\end{equation}

\section{Five-point tree-level numerators}
\label{app:numerators}
The chiral correlator \eqref{e:Lcorr} produces the following sum of ten terms
in the MHV gauge choice of \eqref{e:refmomenta}:
\begin{equation}  \begin{aligned}
-\frac{ (\varepsilon _3 \varepsilon _4 )  (\varepsilon _1 k_4 )  
(\varepsilon _2 k_3 )  (\varepsilon _5 k_3 )}{8 z_{2 3} z_{3 4}} -
\frac{ (\varepsilon _3 \varepsilon _4 )  (\varepsilon _1 k_4 ) 
(\varepsilon _2 k_4 )  (\varepsilon _5 k_3 )}{8 z_{2 4} z_{3 4}} + &
\frac{ (\varepsilon _3 \varepsilon _5 ) (\varepsilon _1 k_3)
(\varepsilon _2 k_3 )  (\varepsilon _4 k_1 )}{8 z_{1 3} z_{2 3}} \\ +
\frac{ (\varepsilon _3 \varepsilon _5 )  (\varepsilon _1 k_3) 
(\varepsilon _2 k_4 )  (\varepsilon _4 k_1 )}{8 z_{1 3} z_{2 4}} +
\frac{ (\varepsilon _3 \varepsilon _5 )  (\varepsilon _1 k_4 ) 
(\varepsilon _2 k_3 )  (\varepsilon _4 k_3 )}{8 z_{2 3} z_{3 4}} - &
\frac{ (\varepsilon _3 \varepsilon _4 ) (\varepsilon _1 k_3) 
(\varepsilon _2 k_3 )  (\varepsilon _5 k_4 )}{8 z_{1 3} z_{2 3} z_{3 4}} \\ -
\frac{ (\varepsilon _3 \varepsilon _5 )  (\varepsilon _1 k_3) 
(\varepsilon _2 k_3 )  (\varepsilon _4 k_3 )}{8 z_{1 3} z_{2 3} z_{3 4}} + 
\frac{ (\varepsilon _3 \varepsilon _5 )  (\varepsilon _1 k_4 ) 
(\varepsilon _2 k_4 )  (\varepsilon _4 k_3 )}{8 z_{2 4} z_{3 4}} - &
\frac{ (\varepsilon _3 \varepsilon _4 ) (\varepsilon _1 k_3) 
(\varepsilon _2 k_4 )  (\varepsilon _5 k_4 )}{8 z_{1 3} z_{2 4} z_{3 4}} \\ - &
\frac{ (\varepsilon _3 \varepsilon _5 )  (\varepsilon _1 k_3) 
(\varepsilon _2 k_4 )  (\varepsilon _4 k_3 )}{8 z_{1 3} z_{2 4} z_{3 4}} \,.
\end{aligned}  \label{e:10terms} \end{equation}
This particular gauge choice killed all double poles. By using the partial
fraction identities of the form \eqref{e:part-frac} to obtain the MSS chiral
block representation.

The Jacobi identities satisfied by the numerators of section~\ref{sec:st-tree} are:
\begin{eqnarray}
 n^{\rm (L/R)}_{3 }  - n^{\rm (L/R)}_{5 } + n^{\rm (L/R)}_{8 } =0  ,\nn\\
 n^{\rm (L/R)}_{3 } - n^{\rm (L/R)}_{1 } + n^{\rm (L/R)}_{12 } =0  ,\nn\\
 n^{\rm (L/R)}_{10 } - n^{\rm (L/R)}_{11 } + n^{\rm (L/R)}_{13 } =0 ,\nn\\
 n^{\rm (L/R)}_{4 } - n^{\rm (L/R)}_{2 } + n^{\rm (L/R)}_{7 } =0  ,\nn\\
 n^{\rm (L/R)}_{4 } - n^{\rm (L/R)}_{1 } + n^{\rm (L/R)}_{15  } =0  ,\label{e:jacobis}\\
 n^{\rm (L/R)}_{10 } - n^{\rm (L/R)}_{9 } + n^{\rm (L/R)}_{15 } =0  ,\nn\\
 n^{\rm (L/R)}_{8 } - n^{\rm (L/R)}_{6 } + n^{\rm (L/R)}_{9 } =0  ,\nn\\
 n^{\rm (L/R)}_{5 } - n^{\rm (L/R)}_{2 } + n^{\rm (L/R)}_{11 } =0  ,\nn\\
 (n^{\rm (L/R)}_{7 } - n^{\rm (L/R)}_{6 } + n^{\rm (L/R)}_{14 } =0)\,.\nn
\end{eqnarray}
where the last one is a linear combination of the others.

The $(2n-5)!!$ color factors are obtained from the six ones of
\eqref{e:5currents} plugged in \eqref{e:BCJnums} and give rise to the expected
result:
\begin{eqnarray}
&& 
c_{1\phantom{0}} =  {f}^{ 1  2 b} {f}^{b  3 c} {f}^{c  4
 5}\,, \hskip 0.8cm 
c_{2\phantom{1}} =  {f}^{ 2  3 b} {f}^{b  4 c} {f}^{c  5
 1}\,, \hskip 0.8cm 
c_{3\phantom{1}} =  {f}^{ 3  4 b} {f}^{b  5 c} {f}^{c  1
 2}\,, \nn \\&&
c_{4\phantom{1}} =  {f}^{ 4  5 b} {f}^{b  1 c} {f}^{c  2
 3}\,, \hskip 0.8cm 
c_{5\phantom{1}} =  {f}^{ 5  1 b} {f}^{b  2 c} {f}^{c  3
 4}\,, \hskip 0.8cm 
c_{6\phantom{1}} =  {f}^{ 1  4 b} {f}^{b  3 c} {f}^{c  2
 5}\,, \nn \\&& 
c_{7\phantom{1}} =  {f}^{ 3  2 b} {f}^{b  5 c} {f}^{c  1
 4}\,, \hskip 0.8cm 
c_{8\phantom{1}} =  {f}^{ 2  5 b} {f}^{b  1 c} {f}^{c  4
 3}\,, \hskip 0.8cm 
c_{9\phantom{1}} =  {f}^{ 1  3 b} {f}^{b  4 c} {f}^{c  2
 5}\,, \nn \\&&
c_{10} =  {f}^{ 4  2 b} {f}^{b  5 c} {f}^{c  1  3}\,,
\hskip 0.8cm  
c_{11} =  {f}^{ 5  1 b} {f}^{b  3 c} {f}^{c  4  2}\,,
\hskip 0.8cm  
c_{12} =  {f}^{ 1  2 b} {f}^{b  4 c} {f}^{c  3  5}\,,
\nn \\\ &&
c_{13} =  {f}^{ 3  5 b} {f}^{b  1 c} {f}^{c  2  4}\,,
\hskip 0.8cm  
c_{14} =  {f}^{ 1  4 b} {f}^{b  2 c} {f}^{c  3  5}\,,
\hskip 0.8cm  
c_{15} =  {f}^{ 1  3 b} {f}^{b  2 c} {f}^{c  4  5}\,.
\label{e:5color}
\end{eqnarray}

\section{Integrating the triangles}
\label{app:triangles}

The BCJ triangle numerators~\eqref{trianglesfinal} are linear in the loop momentum,
so if we apply to them the exponentiation procedure of section~\ref{sec:loop-time},
we get the following terms
\begin{equation}
      n_{\rm tri}(\tell+K) \propto B_{\mu} (\tell^{\mu} + K^{\mu}) + C  \,,
\end{equation}
where $ K = - \sum u_i k_i $. The linear term linear integrates to zero by parity
and the constant term $BK+C$ vanishes for each triangle numerator.
For example, for the numerator~\eqref{triangleexample},
\begin{equation}
      B^{\mu} = - s k_3^{\mu} + t k_1^{\mu} - u k_2^{\mu}
                + \frac{4iu}{s} k_{1 \mu_1} k_{2 \mu_2} k_{3 \mu_3}
                                 \epsilon^{\mu_1 \mu_2 \mu_3 \mu} \,, ~~~~~
      C = s u \,.
\end{equation}
So it can be easily checked that
\begin{equation}
      B_{\mu} K^{\mu} = u_1 s u - u_4 s u \,.
\end{equation}
Taking into account that $ u_4 = 1 $ and that the particular triangle~\eqref{triangleexample}
is obtained from the worldline box parametrization by setting $ u_1 = 0 $,
we indeed obtain $BK = -su = -C$.

Moreover, in the gravity amplitude, the triangle numerators squared become simply:
\begin{equation}
      n_{\rm tri}^2(\tell+K) \propto B_{\mu} B_{\nu} \tell^{\mu} \tell^{\nu} \,.
\end{equation}
The standard tensor reduction transforms $\tell^{\mu} \tell^{\nu}$
to $\ell^2\eta^{\mu\nu}/4$, which is known to induce a dimension shift \cite{Bern:1995db}
from $d=4-2\epsilon$ to $d=6-2\epsilon$.
As a result, in the double copy construction the BCJ triangles produce
six-dimensional scalar triangle integrals~\eqref{tri1m6d}
with the coefficients~\eqref{e:trianglecoeff}.

\section{Explicit expression of $\delta W_3$}
\label{app:dW}

In section~\ref{sec:comp-gauge}, we expressed $\delta W_3$ in terms of $\dot
G$'s:
\begin{equation} \begin{aligned}
   \delta W_3\, = ~\frac{1}{2}\,
      \Big( (1 + 2\As - 2A_1 - 2A_2) ({\dot G}_{14}^2 - {\dot G}_{23}^2)
          - 2 (1 + A_1) {\dot G}_{12}^2 - 2 (1 - A_1) {\dot G}_{34}^2 \:\, & \\
          - 2 A_3 ({\dot G}_{13} - {\dot G}_{14} - {\dot G}_{23} + {\dot G}_{24})
                  ({\dot G}_{13} - {\dot G}_{14} + {\dot G}_{23} - {\dot G}_{24}
                                                               + 2 {\dot G}_{34}) & \\
          + 2 (1 - 2 \As + 2 A_1 + A_2)
              ({\dot G}_{12} {\dot G}_{14} - {\dot G}_{12} {\dot G}_{24} 
                                           + {\dot G}_{14} {\dot G}_{24}) & \\
          - 2 (2 + 2 \As - 2 A_1 - A_2)
              ({\dot G}_{23} {\dot G}_{34} - {\dot G}_{24} {\dot G}_{34}
                                           - {\dot G}_{23} {\dot G}_{24}) & \\
          + 2 (1 - A_2)({\dot G}_{13} {\dot G}_{34} - {\dot G}_{14} {\dot G}_{34}
                                                    - {\dot G}_{13} {\dot G}_{14}) & \\
          - 2 A_2 ({\dot G}_{12} {\dot G}_{13} - {\dot G}_{12} {\dot G}_{23}
                                               + {\dot G}_{13} {\dot G}_{23}) &
      \Big) \\
    - \,\frac{tu}{s^2}\, \big( 1 - \As + \Ae \big)
      \Big(  {\dot G}_{14}^2 - {\dot G}_{23}^2 - 2{\dot G}_{12} {\dot G}_{14}
          + 2{\dot G}_{12} {\dot G}_{24} - 2{\dot G}_{14} {\dot G}_{24} & \\
          + 2{\dot G}_{23} {\dot G}_{24} - 2{\dot G}_{23} {\dot G}_{34}
          + 2{\dot G}_{24} {\dot G}_{34} &
      \Big) \,,
      \label{e:deltaWgen}
\end{aligned} \end{equation}
where $\As$ and $\Ae$ are the free parameters of the BCJ ansatz,
and $A_1$, $A_2$ and $A_3$ are those from matching to a string-inspired ansatz.

\section{Trick to rewrite the square-correcting terms}
\label{app:sqt}
In this appendix, we use a trick to partly rewrite the square-correcting
terms~\eqref{e:nboxsqt} as string-based quantities. This section is mostly
provided here for the interesting identity~\eqref{e:ddAA} which relates the BCJ
triangles to the quadratic part of the box numerators.

First, we introduce a new element in the reduction technique.
Recall that factors of $1/T^k$ modify the overall factor $1/T^{d/2-(n-1)}$
and thus act as dimension shifts $d \to d+2k$.
Therefore, ${(2A_{\mu\nu}K^\nu+B_\mu)^2/(2T)}$
is the numerator of a six-dimensional worldline box. 

However, we choose to treat the $1/T^2$ differently. 
Since $A_{\mu\nu}$ does not depend on the ordering, we can rewrite the $1/T^2$
square-correcting term as a full worldline integral 
\begin{equation}
      \frac{i}{(4\pi)^{\frac{d}{2}}}
      \frac{\eta^{\mu(\nu} \eta^{\rho\sigma)} A_{\mu\nu} A_{\rho\sigma}}{4}
      \int_0^\infty \!\!\!{\d T\over T^{\frac{d}{2}-3}}
      \int \d^3 u \frac{e^{-T Q}}{T^2} \,,
\label{e:Aterm}
\end{equation}
where the proper-time domain in $u_i$ contains all three inequivalent box
orderings.
Now let us consider the second derivative of the worldline propagator
\begin{equation}
      \ddot G_{ij} = \frac{2}{T}\left(\delta(u_{ij}) -1\right) \,,
\end{equation}
to obtain a useful identity valid for any $i,j,k,l$:
\begin{equation}
      \frac{1}{T^2} = \frac{1}{4} \ddot G_{ij} \ddot G_{kl}
    + \frac{1}{T^2} \left(\delta(u_{ij})+\delta(u_{kl})\right)
    - \frac{1}{T^2} \delta(u_{ij})\delta(u_{kl}) \,.
\label{e:1T2}
\end{equation}
The factors of $1/T^2$ combine with delta-functions and thus properly change the
number of external legs and dimensions, such that from the right-hand side of
\eqref{e:1T2},
we can read off the following integrals: a four-dimensional worldline box with
numerator $\ddot G_{ij}
\ddot G_{kl}$,
two six-dimensional scalar triangles and
a four-dimensional scalar bubble.
Since we are free to choose indices $i,j,k,l$,
we can as well use a linear combination of the three several choices,
as long as we correctly average the sum.
For instance, we can now create $s$-, $t$- and $u$-channel six-dimensional
scalar triangles
(along with four-dimensional scalar bubbles),
if we choose $(i,j,k,l)\in\{(1,2,3,4),(1,4,2,3),(1,3,2,4)\}$
and sum over them with coefficients $\lambda_s, \lambda_t$ and $\lambda_u$:
\begin{equation} \begin{aligned}
      \int_0^\infty \!\!\!{\d T\over T^{\frac{d}{2}-3}}
      \int \d^3 u \frac{e^{-T Q}}{T^2}
  = & \bigg[\, \frac{1}{4}
      \int_0^\infty \!\!\!{\d T\over T^{\frac{d}{2}-3}}
      \int \d^3 u \left(
      \lambda_s \ddot G_{12} \ddot G_{34}
    + \lambda_t \ddot G_{14} \ddot G_{23}
    + \lambda_u \ddot G_{13} \ddot G_{24} \right) e^{-T Q} \\
 & +  \, 2 \sum_{c=s,t,u} \lambda_c I_3^{6d}(c)
    - \frac{1}{2} \sum_{c=s,t,u} \lambda_c I_2^{4d}(c) \bigg]\frac 1
{\sum_{c=s,t,u} \lambda_c}\,,
\end{aligned} \label{e:redT^-2} \end{equation}
This expression is written at the integrated level,
in order to be completely explicit with the subtle normalizations. 
In particular, we took into account that the scalar triangles coming from
$\delta(u_{12})$ and $\delta(u_{34})$ depend only on $s$ and have equal
integrated contributions.
We summed over the three orderings and used the fact
that $\delta(u_{ij})$ generate factors of $1/2$ due to always acting on the
border of their proper-time domains.

To sum up, the term ${\eta^{\mu(\nu}\eta^{\rho\sigma)}
A_{\mu\nu}A_{\rho\sigma}/(4 T^2)}$
in \eqref{e:Aterm} produces the three following scalar triangles:
\begin{equation}
      \frac{i}{(4\pi)^{\frac{d}{2}}}
      \frac{ \eta^{\mu(\nu}\eta^{\rho\sigma)} A_{\mu\nu}A_{\rho\sigma} }
           { 4 \sum_{c=s,t,u} \lambda_c }
      \sum_{c=s,t,u} (2 \lambda_c) I_3^{6d}(c) \,,
\label{e:trianglesfromboxes}
\end{equation}
in addition to the four-dimensional non-scalar boxes and scalar bubbles.

At this point, let us recall that the BCJ gravity amplitude contains
triangles that integrate to six-dimensional scalar triangles:
\begin{equation} \begin{aligned}
      \sum \int \frac{\d^d \ell}{(2\pi)^d}
      \frac{n_{\rm tri}^2(\ell)}{D_{\rm tri}(\ell)}
      = - \frac{i}{(4\pi)^{\frac{d}{2}}}
           \sum_{c=s,t,u} (\lambda_{c_1}+\lambda_{c_2}) I_3^{6d}(c) \,,
\label{e:gravtriangles}
\end{aligned} \end{equation}
with coefficients:
\begin{equation} \begin{aligned}
      \lambda_{s_1} & = \frac{( (1-\As) s^2 + 2 (1-\As+\Ae) tu )^2}
                          {8 s^4 t u} \,, \\
      \lambda_{s_2} & = \frac{( (1-\As) s^2 - 2 (1-\As+\Ae) tu )^2}
                          {8 s^4 t u} \,, \\
      \lambda_{t_1} & = \lambda_{t_2} = \frac{s^2 - 4 u^2}{8 s^3 u} \,, \\
      \lambda_{u_1} & = \lambda_{u_2} = \frac{s^2 - 4 t^2}{8 s^3 t} \,.
\end{aligned} \label{e:trianglecoeff} \end{equation}

Therefore, we can try to match the actual coefficient of the BCJ triangles in
\eqref{e:gravtriangles} by choosing different fudge factors $\lambda_s,
\lambda_t$ and $\lambda_u$ in \eqref{e:trianglesfromboxes} for the triangles
generated by the reduction of the $1/T^2$ term as follows:
\eqref{e:gravtriangles}:
\begin{equation}
      \lambda_c = \lambda_{c_1} + \lambda_{c_2} \,,\qquad  c=s,t,u\,.
\label{e:lambdas}
\end{equation}
This choice implies that for any values of $\As$ and $\Ae$
\begin{equation}
      {\eta^{\mu(\nu}\eta^{\rho\sigma)} A_{\mu\nu} A_{\rho\sigma}
      \over 4 \sum_{c=s,t,u} \lambda_c} = 1 \,.
\label{e:ddAA}
\end{equation}
This lets us carefully relate the scalar-triangle contributions
\eqref{e:trianglesfromboxes} coming from the square-correcting
terms~\eqref{e:Aterm}
to be equal to the $(-2)$ times the BCJ triangles squared.\footnote{Recall that
the overall normalization of $\lambda_c$'s was irrelevant
in eq.~\eqref{e:redT^-2}, so the factor $(-2)$ is fixed
and prevents us from completely eliminating the triangles.}

This seeming coincidence deserves a few comments.
We defined $A_{\mu\nu}$ as the coefficient of $\ell^{\mu}\ell^{\nu}$
in the BCJ box numerators \eqref{boxesfinal},
but in principle, we know that the boxes could have been made scalar
in the scalar integral basis, as in \eqref{N1chiral}.
To comply with the kinematic Jacobi identities,
the BCJ color-kinematics duality reintroduces $\ell^{2}$ into the boxes
by shuffling them with the scalar triangles and bubbles.
In our final BCJ construction, we set bubble numerators to zero,
so the information that was inside the original scalar triangles and bubbles
was equally encoded in the dependence of the BCJ box and triangle numerators
on the loop momentum.
This is why the coincidence between the $A_{\mu\nu}$ and $\lambda_c$
is not miraculous.

Finally, we can rewrite eq.~\eqref{e:W2-simp1} using our trick:
\begin{equation} \begin{aligned}
   \int \frac{1}{2} (W_2 & + \delta W_2) = \int \bigg\{\!
      - \sum \langle n_{\rm tri}^2 \rangle
      - \sum_{\mathfrak{S}} \frac{(2A_{\mu\nu} K^\nu + B^{(\mathfrak{S})}_\mu)^2}{2T} \\
    & + \frac{1}{4} \left( \lambda_s \ddot G_{12} \ddot G_{34}
                      \!+\!\lambda_t \ddot G_{14} \ddot G_{23}
                      \!+\!\lambda_u \ddot G_{13} \ddot G_{24} \right)
      - \frac{1}{T^2} \left( \lambda_s \delta_{12} \delta_{34}
                        \!+\!\lambda_t \delta_{14} \delta_{23}
                        \!+\!\lambda_u \delta_{13} \delta_{24} \right) \!\bigg\} \,.
\label{e:W2-simp2}
\end{aligned} \end{equation}

We could not apply the same trick to the $1/T$ square-correcting terms
because they do not seem to have a nice string-theoretic interpretation
with respect to the ``string-ancestor-gives-no-triangles''criterion.
More precisely, we expressed it as a worldline polynomial by the same ansatz
method that we used to determine the expression of $\delta W_3$, and observed
explicitly that this term does not satisfy this criterion, i.e. it creates
triangles in the field theory limit.
Moreover, we checked the non-trivial fact that the coefficients of these
triangles cannot be made equal to these of the BCJ triangles.

\bibliographystyle{JHEP}
\bibliography{bcj}

\nocite{*}

\end{document}